\documentclass[aps,pra,twocolumn,longbibliography]{revtex4-1}

\usepackage{amsmath,amssymb}
\usepackage{mathrsfs}
\usepackage{graphicx}
\usepackage{physics}
\usepackage{xcolor}
\usepackage{dsfont}
\usepackage[hidelinks]{hyperref}
\usepackage[normalem]{ulem}
\usepackage[utf8]{inputenc}
\usepackage[T1]{fontenc}

\hypersetup{colorlinks=true, linkcolor=black, citecolor=black, urlcolor=blue}

\begin{document}

\title{Non-equilibrium many-body dynamics in supersymmetric quenching}
\author{Christopher Campbell}
\email[Corresponding author: ]{christopher.campbell@oist.jp}
\author{Thom\'as Fogarty}
\affiliation{Quantum Systems Unit, Okinawa Institute of Science and Technology Graduate University, Onna, Okinawa 904-0495, Japan} 
\author{Thomas Busch}
\affiliation{Quantum Systems Unit, Okinawa Institute of Science and Technology Graduate University, Onna, Okinawa 904-0495, Japan} 
\date{\today}

\begin{abstract}
 We study the dynamics induced by quenching an ultracold quantum many-body system between two supersymmetric Hamiltonians. Such a quench can be created by carefully changing the external trapping potential and leads to a situation where the eigenspectra before and after the quench are nearly identical. We show that the dynamics originating from this can be conveniently described using knowledge about the initial state only and apply this insight to the specific example of a fermionic gas that is initially trapped in an infinite box potential. Quenching to different, higher order supersymmetric partners potentials we observe the appearance of many-body revivals in the survival probability and show that some of these are robust at finite temperatures. This is in contrast to the well known Talbot effect, which is the standard example for quenching into a system with a quadratic spectrum. 
\end{abstract}

\maketitle

\section{Introduction}

The ongoing advances in the control and measurement of ultracold quantum gases has established these systems as highly promising laboratories to study many-body quantum dynamics \cite{Mistakidis:22}. Of particular interest in this area is the evolution of a system after a quench of its internal or external parameters and the question of how the large number of degrees of freedom leads to equilibration and thermalization over time \cite{Eisert:15}. To study these, a system is usually considered to be prepared in a pure state, typically the ground state of a potential, and a sudden change in the Hamiltonian parameters or the addition of an external perturbation is used to excite it. Many ways to do this have been proposed and studied in recent years, ranging from from giving a momentum kick to a quantum gas \cite{Kinoshita:06}, changing the trapping potential \cite{Campo_2006,Collura13,Kazuma:19}, adjusting the interparticle interaction strengths \cite{Mistakidis_2014,GarciaMarch:16}, or introducing an impurity \cite{Knap:12,Campbell:14}. 

Once the system is excited, the subsequent non-equilibrium dynamics is often rather complex. However, over the past decade a number of quantities and tools have been developed that allow one to capture typical behaviour and identify underlying mechanisms, such as the Loschmidt echo \cite{Tim:Quench:2016,MOSSY:ORTHO:2020}, the work probability distribution \cite{fusco2014,Campbell16} and the amount of information scrambling \cite{Chenu2018,Chenu2019workstatistics,Mikkelsen22}.  Most quenches that rely on a change of the Hamiltonian parameters induce the non-equilibrium dynamics via a change in the eigenspectrum of the system. In this case states that were previously stationary become superpositions of eigenstates, which leads to dynamics that are strongly affected by the relation between the initial and the final spectrum. Interesting examples here are quenches in shape invariant potentials \cite{Deffner14}, such as harmonic traps \cite{Minguzzi05,Rohringer2015,Collura18}, and box potentials \cite{Friesch_2000,Campo2012,Vicari19} where the initial and final spectrum are related through a scaling factor. These systems can exhibit perfect revivals of the initial state for both single particle and many-body systems, similar to the well-known Talbot effect in infinite box potentials which has also been thoroughly studied for both bosonic and fermionic cold atoms \cite{Grochowski20,Revivals:StoM:2021}. It is also worth noting that the recurrences of many-body states are related to the concepts of quantum scars, which signal weak ergodicity breaking after a global quench \cite{SUSY_ergodicity:2020,Enhanced_scar:2020,Erg_breaking:2018}. 

In this work we study a specific type of quench that leaves the eigenspectrum unperturbed, but removes a fixed number of states from below. This situation appears when quenching between two different Hamiltonians that are related through a supersymmetric algebra, which results in the eigenspectra between the two potentials being near identical \cite{COOPER:SUSYBIBLE:1997,onedDSUSY:1985}. 
The supersymmetric relationship also provides a number of tools that can be used to transform a wavefunction between the two partner potentials, namely the supersymmetric operators and the superpotential. We will show that these allow us to recast the non-equilibrium dynamics of the post quench system purely in terms of the the initial Hamiltonian and its supersymmetric operators. While this has many advantages from a practical perspective, it also allows us to highlight and understand the unique dynamics that emerge in supersymmetric quenches. This is therefore another example where supersymmetric setups can be used in quantum systems for controlled dynamics, similar to approaches developed for the ground state preparation of ultracold atoms arrays \cite{Luo2021} or mode filtering  in optical setups \cite{Heinrich2014}.

In the following we will first briefly review the  concepts of supersymmetric quantum mechanics relevant to our work and introduce the main quantities we will use to describe a quench between two supersymmetric Hamiltonians. We then study the ensuing dynamics at zero and at finite temperatures and in particular the appearance and stability of full revivals in the survival probability.  These results are also compared to the related Talbot effect.

\begin{figure}[tb]
    \centering
    \includegraphics[width =\linewidth]{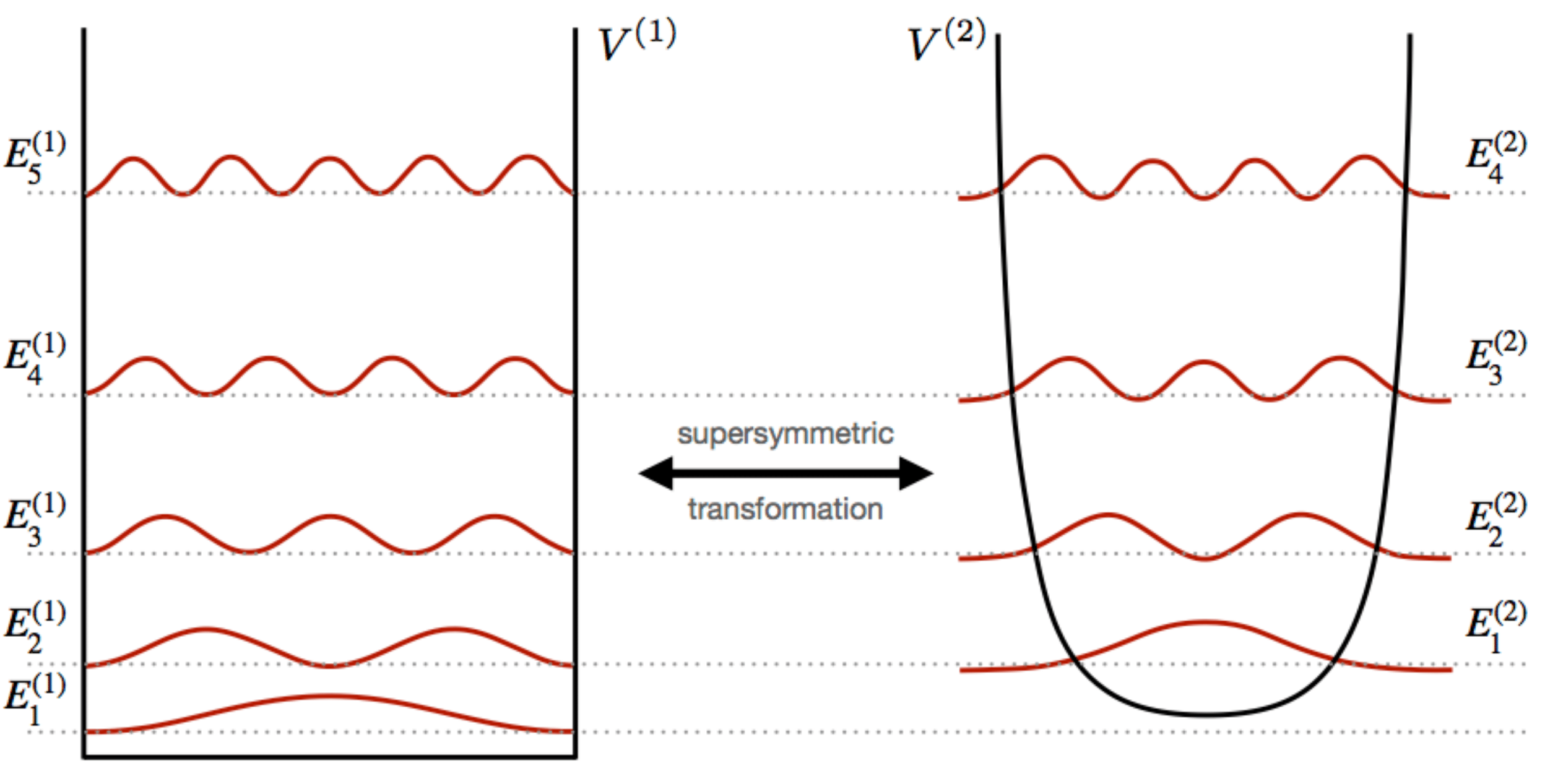}
    \caption{Infinite box potential, $V^{(1)}$, and its first  supersymmetric partner, $V^{(2)}=\frac{\pi^2}{2L^2} \left( 2\sec^2(\frac{\pi x}{L})-1 \right)$. The spectrum is indicated by the dashed lines and the densities of the eigenfunctions, $|\psi_n^{(1)}|^2$ and $|\psi_m^{(2)}|^2$, are given by the red curves.}
    \label{fig:SUSYSchematic}
\end{figure}

\subsection{Supersymmetric Hamiltonians}

A supersymmetric algebra relates two Hamiltonians that can be written in a factorized manner as $H^{(1)} = A_1^\dagger A_1$ and  $H^{(2)} = A_1 A_1^\dagger$,  where \cite{Hull:factorization:1951,COOPER:SUSYBIBLE:1997},
\begin{align}
    A_1 =& \frac{\hbar}{\sqrt{2m}}\frac{d}{dx} + \mathcal{W}(x),\\
    A_1^\dagger =& -\frac{\hbar}{\sqrt{2m}}\frac{d}{dx} + \mathcal{W}(x).
    \label{eq:SUSY_ops}
\end{align}
The function $\mathcal{W}(x)$ is known as the superpotential and writing the Hamiltonians in the explicit form
\begin{eqnarray}
    H^{(1)} &= -\frac{\hbar^2}{2m}\frac{d^2}{dx^2} - \frac{\hbar}{\sqrt{2m}}\mathcal{W}'(x)+ \mathcal{W}(x)^2,\\
    H^{(2)} &= -\frac{\hbar^2}{2m}\frac{d^2}{dx^2} + \frac{\hbar}{\sqrt{2m}}\mathcal{W}'(x)+ \mathcal{W}(x)^2,
    \label{eq:H2}
\end{eqnarray}
allows one to find two supersymmetric potentials as
\begin{align}
    V^{(1)}(x)&= \mathcal{W}(x)^2-\frac{\hbar}{\sqrt{2m}}\mathcal{W}'(x),\\ 
    V^{(2)}(x)&= \mathcal{W}(x)^2+\frac{\hbar}{\sqrt{2m}}\mathcal{W}'(x).
\end{align}
The superpotential itself can be determined in a number of ways \cite{COOPER:SUSYBIBLE:1997}, with a straightforward one being given by the logarithmic derivative of the ground state wavefunction of a given $H^{(1)}$ as
\begin{equation}\label{eq:superpot}
    \mathcal{W}(x) = -\frac{\hbar}{\sqrt{2m}}\partial_x\ln\left(\psi^{(1)}_0\right).
\end{equation}  
An important property of supersymmetric Hamiltonians is that they possess near identical eigenspectra, with the one for the second Hamiltonian shifted upwards by one quantum number, i.e.~$E_{m}^{(2)}=E_{m+1}^{(1)}$ (see Fig.~\ref{fig:SUSYSchematic} for the example of an infinite well \cite{infinite_well_SUSY:2018}). 

From the factorized form of the Hamiltonians one can immediately see that an intertwining relationship between them exists 
\begin{align}
    \label{eq:first_intertwine}
    A_1H^{(1)} &= H^{(2)}A_1, \\
    H^{(1)} A_1^\dagger &= A_1^\dagger H^{(2)},
\end{align}
which can be used to connect the respective eigenfunctions as \cite{Cooper_1989:SUSY_Transform} 
\begin{equation}
    \label{eq:WFIntertwine}
    \psi^{(2)}_m = \frac{A_1}{\sqrt{\Delta E_m}}\psi^{(1)}_{m+1},
\end{equation}
where $\Delta E_m = E^{(2)}_{m} - E^{(1)}_0$. The operators $A$ and $A^\dagger$ therefore map even wavefunctions into odd ones.

The intertwining property between the two Hamiltonians also extends to the time propagation operator $U(t) = e^{-\frac{i}{\hbar}Ht}$ \cite{SUSY:Intertwine:2017}. This can be easily seen by expanding the exponential in a power series as
\begin{equation}
    e^{-\frac{i}{\hbar}H^{(2)t}} = \sum^\infty_{k=0}\frac{1}{k!}\left(-\frac{i}{\hbar}H^{(2)}t\right)^k,
\end{equation}
and noting that all terms in the sum are of the form $(A_1A_1^\dagger)^k$. Applying an annihilation operator from the right then leads to $A_1(H^{(1)})^k$ and therefore to
\begin{equation}
    \label{eq:TEIntertwine}
    A_1U^{(1)}(t) = U^{(2)}(t)A_1 .
\end{equation}
Iterating on the supersymmetric relation between $H^{(1)}$ and $H^{(2)}$, one can create a hierarchy of supersymmetric partner potentials by finding a factorization of $H^{(2)}=A_2^{\dagger} A_2 $ \cite{Hierarchy:2004}. The resulting partner Hamiltonian, $H^{(3)}=A_2 A_2^{\dagger}$, has a spectrum that is degenerate with the ones of $H^{(1)}$ and $H^{(2)}$, however it is missing the ground state energies of each. The transformations between higher order Hamiltonians require their own set of operators, however the connection between the eigenfunctions of the, say, first and third supersymmetric Hamiltonian is simply given by
\begin{equation}
    \psi^{(1)}_{m+2} = \frac{1}{\sqrt{E^{(3)}_m-E^{(1)}_0}}\frac{1}{\sqrt{E^{(3)}_m-E^{(2)}_0}}A^\dagger_1 A^\dagger_2\psi^{(3)}_m\;.
\end{equation}

\subsection{Survival Probability and Dynamical Overlap}

In the following we will consider an effectively one-dimensional, spin-polarized Fermi gas of $N$ neutral atoms at low temperatures \cite{Liao:10}. The gas is assumed to be confined in a potential that can be changed time-dependently with a high degree of flexibility, which can be experimentally realised in the recently emerging  setups that use SLMs \cite{Henderson_2009,Paint_potential:SLM:2018}. The many-body states $\mathcal{H}^{(\alpha)}|\Psi_n^{(\alpha)}\rangle=\mathcal{E}_n^{(\alpha)}|\Psi_n^{(\alpha)}\rangle$ are determined by the Hamiltonian 
\begin{equation}
    \mathcal{H}^{(\alpha)} = \sum^N_{j =1}\left[-\frac{\hbar^2}{2m}\nabla^2_j + V^{(\alpha)}(x_j)\right],
\end{equation}
where $\alpha$ represents the order of the supersymmetric hierarchy and $\alpha=1$ is the original potential we quench from. The initial state at inverse temperature $\beta=(k_B T)^{-1}$ is given by $\rho^{(1)}=\sum_n \frac{1}{Z_0}e^{\beta(\mathcal{E}_n^{(1)}-N\mu)}\vert \Psi_n^{(1)} \rangle \langle \Psi_n^{(1)} \vert$, where $\mu$ is the chemical potential of the system, $k_B$ is the Boltzmann constant and $Z_0$ is the grand canonical partition function. 

The non-equilibrium dynamics induced by a quench is generally rather complex and a good quantity to characterise it is the survival probability, which describes the squared overlap of the initial state with the time-evolved one. For a quench from $\mathcal{H}^{(1)}$ to $\mathcal{H}^{(2)}$ it is given by $F(t)=|\mathcal{O}(t)|^2=\left|\Tr\left[e^{\frac{i}{\hbar}\mathcal{H}^{(2)}t}e^{-\frac{i}{\hbar}\mathcal{H}^{(1)}t} \rho^{(1)}\right]\right|^2$. The calculation of the dynamical overlap $\mathcal{O}(t)$ can be eased considerably by rewriting the many-body states $|\Psi_n^{(\alpha)}\rangle$ as a Slater determinant of single particle states $H^{(\alpha)}|\psi^{(\alpha)}_k\rangle = E^{(\alpha)}_k |\psi^{(\alpha)}_k\rangle$, which gives \cite{Levitov:2005,Fermi_edge_res:2005,Demler:ORTHO:2012,KnapDemler:2018}
\begin{equation}
    \label{eq:survival_probability}
    \mathcal{O}(t)=\det\left[1-\hat{n} + \hat{n}e^{\frac{i}{\hbar}H^{(2)}t}e^{-\frac{i}{\hbar}H^{(1)}t}\right],
\end{equation}
where $\hat{n} = (e^{\beta(H^{(1)}-\mu)}+1)^{-1}$ is the Fermi--Dirac distribution and $H^{(\alpha)}=-\frac{\hbar^2}{2m}\nabla^2 + V^{(\alpha)}(x)$ is the corresponding single particle Hamiltonian. 

At zero temperature the initial many-body ground state is a perfect Fermi sea described by
\begin{equation}
    \Psi_0^{(1)}\left(x_1,x_2,...,x_N\right) = \frac{1}{\sqrt{N!}}\det_{k,j=1}^N[\psi^{(1)}_k(x_j)],
    \label{eq:slater}
\end{equation}
which allows one to write the survival probability in a simpler way as \cite{Campo:decay_SP:2011,Mossy:ORTHO:2011,Demler:ORTHO:2012}
\begin{equation}
    F_{T=0}(t) = 
    \left|\bra{\Psi^{(1)}_0}e^{\frac{i}{\hbar}\mathcal{H}^{(2)}t}e^{-\frac{i}{\hbar}\mathcal{H}^{(1)}t}\ket{\Psi^{(1)}_0}\right|^2.
\end{equation}
This can be simplified further by introducing the dynamical overlap probabilities of the single particle states as
\begin{align}
    \label{eq:SingleParticleOverlap}
    \mathcal{O}_{kl}(t)=&\bra{\psi^{(1)}_k}e^{\frac{i}{\hbar}H^{(2)}t}e^{-\frac{i}{\hbar}H^{(1)}t} \ket{\psi^{(1)}_l}\nonumber\\
    =&\sum_{m=1}^\infty\bra{\psi^{(1)}_k}\ket{\psi^{(2)}_m}\bra{\psi^{(2)}_m}\ket{\psi^{(1)}_l}e^{-\frac{i}{\hbar}\left(E_l^{(1)}-E_{m}^{(2)}\right)t},
\end{align}
leading to
\begin{equation}
    \label{eq:survival_probability_T0}
    F_{T=0}(t) =\big|\det\left[\mathcal{O}_{kl}(t)\right]\big|^2\;,
\end{equation}
which is equivalent to Eq.~\eqref{eq:survival_probability} at $T=0$.

It is interesting to note that the time-dependence of the survival probability can be fully predicted from the knowledge of the initial single particle Hamiltonian since, by using the transformation between the two sets of eigenstates of the two supersymmetric Hamiltonians given in Eq.~\eqref{eq:WFIntertwine}, the single particle overlaps can be expressed purely in terms of the eigenstates of the initial Hamiltonian as
\begin{align}
    \label{eq:overlap_matrix}
     \mathcal{O}_{kl}=
     \sum_{m=1}^\infty\bra{\psi^{(1)}_k}A\ket{\psi^{(1)}_{m+1}}&\bra{\psi^{(1)}_{m+1}}A^\dagger\ket{\psi^{(1)}_{l}}\nonumber\\
     &\times\frac{e^{-\frac{i}{\hbar}\left(E_l^{(1)}-E_{m+1}^{(1)}\right)t}}{\Delta E_m}\;.
\end{align}

The survival probability indicates how distinguishable the non-equilibrium state after the quench $|\Psi(t)\rangle$ is from the state at $t=0$. The initial decay of the survival probability is proportional to the strength of the quench, i.e. how different $H^{(2)}$ is from $H^{(1)}$, and is related to the amount of irreversible excitations created \cite{GarciaMarch:16}. Complete revivals of the survival probability, $F(t)=1$, indicate the recurrence of the initial state and are usually the result of finite size effects stemming from the trapping potential \cite{GarciaMarch:16,Grochowski20,Revivals:StoM:2021}. 

\section{Quench at Zero Temperature}

To explore quenching between supersymmetric potentials we start by assuming that the $N$ particle system is initially confined in an infinite square well of width $L$ 
\begin{equation}
  V^{(1)}(x)=\begin{cases}
    \infty, & \text{if}\; |x|>\frac{L}{2}\;,\\
    0, & \text{if}\; |x|<\frac{L}{2}\;.
  \end{cases}
\end{equation}
For this system the single particle eigenstates are well known and given by
\begin{equation}
  \psi^{(1)}_n(x)=\begin{cases}
    \sqrt{\frac{2}{L}}\text{sin}(\frac{n x\pi}{L}), & \; n = \text{even;},\\
    \sqrt{\frac{2}{L}}\text{cos}\left(\frac{n x\pi}{L}\right), & \; n = \text{odd},
  \end{cases}
\end{equation}
with the corresponding eigenenergies being $E^{(1)}_n=n^2\pi^2\hbar^2/(2mL^2)$. The infinite box potential is a common system for which supersymmetric partners have been studied extensively \cite{infinite_well_SUSY:2018}, and, indeed, the superpotential for $V^{(1)}(x)$ and its partner potential $V^{(2)}(x)$ can be calculated as $\mathcal{W}^{(2)}(x) = \frac{\pi}{\sqrt{2}L}\tan \left(\frac{\pi x}{L}\right)$ where the superscript $(2)$ references the target potential. The exact eigenfunctions of the supersymmetric Hamiltonian $H^{(2)}$ are known and the expression for the general superpotential for higher partners is given by \cite{infinite_well_SUSY:2018,Sukumar_1985}
\begin{equation}\label{eq:inf_superpot}
    \mathcal{W}^{(\alpha)} = (\alpha - 1)\frac{\pi}{\sqrt{2}L}\text{tan}\left(\frac{x\pi}{L}\right)\;.
\end{equation}

\begin{figure}[tb]
    \centering
    \includegraphics[width =\linewidth]{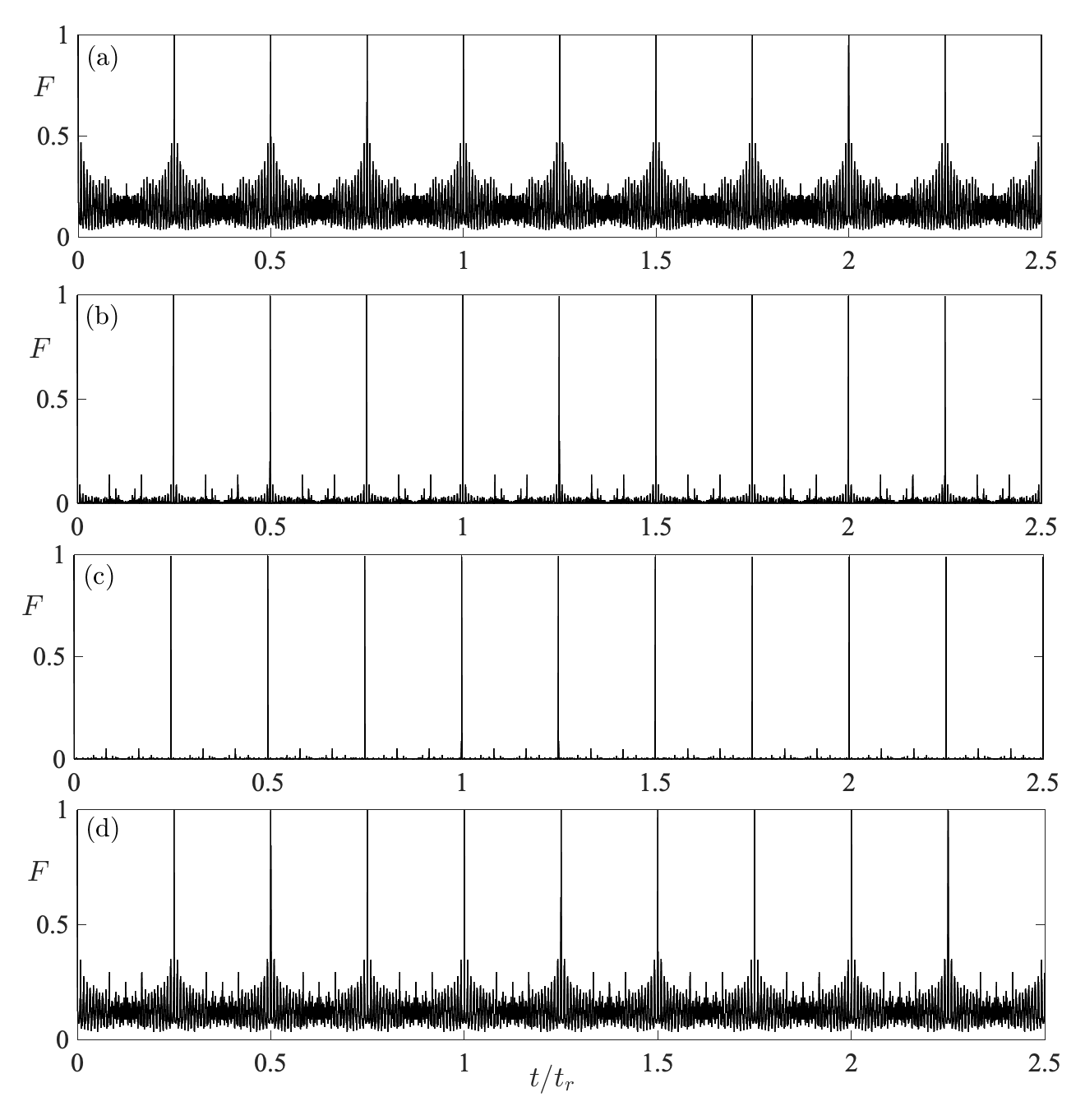}
    \caption{Survival probabilities for different quenches of a system of $N=30$ fermions at $T=0$ starting with an infinite box of width $L=4$.  In panels (a-c) the final potentials are the supersymmetric partner potentials $V^{(2)}$, $V^{(3)}$ and $V^{(4)}$, respectively. In panel (d) the quench is from a box of length $L'=3.9$ to one of length $L=4$. Note that the parameters for the quench in (d) were chosen to compare to the dynamics observed in (a).}
    \label{fig:SurvivalT0}
\end{figure}

In the following we focus on the dynamics of a quench from an initial box potential $V^{(1)}(x)$ directly to a higher order supersymmetric potential $V^{(2)}(x)$, $V^{(3)}(x)$ or $V^{(4)}(x)$. The initial state at $T=0$ is the groundstate $|\Psi_0^{(1)}\rangle$ of the Hamiltonian $H^{(1)}$ and, for a gas of $N=30$ fermions, the survival probability is shown in Figs.~\ref{fig:SurvivalT0}(a-c). One can see that in all cases the survival probability decays quickly immediately after the quench, but regular complete recurrences of the initial state ($F(t)=1$) appear as well. In fact, these occur at integer multiples of $t_r/4$, where the standard, state-independent revival time for a box is given by \cite{Revivals:1997,Revivals:infinite_explained:2000}
\begin{equation}\label{revival_time}
        t_r = \frac{4mL^2}{\pi\hbar} = \frac{2\pi\hbar}{E^{(1)}_1},
\end{equation}
and where $E^{(1)}_1$ is the ground state energy of the initial Hamiltonian.

The occurrence of the revivals at multiples of $t_r/4$ can be easily understood by realising that at these times the phase evolution of all single particle eigenstates is independent of $n$ \cite{Revivals_revisited,Bluhm:WF_Revival:1996,Revivals:1997,Revivals:infinite_explained:2000,ROBINETT:Revivals:2004,Revivals:1997}. This can be shown by separating the evolving single particle eigenstates into their even and odd sub-spaces, which have the associated energies
\begin{equation}
    E_n^e = E_1(2n+1)^2\quad\text{and}\quad
    E_n^o = 4E_1(n+1)^2\;,
\end{equation}
for $n = 0,1,2,3\dots$. The time evolution is then given by
\begin{align}
   e^{-\frac{i}{\hbar}E^e_n t} &= 
   e^{-i\phi^e_n(t)}=e^{-\frac{2\pi i(2n+1)^2}{t_{r}}t}\;,\\
   e^{-\frac{i}{\hbar}E^o_n t} &= 
   e^{-i\phi^o_n(t)}=e^{-\frac{8\pi i(n+1)^2}{t_{r}}t}\;.
\end{align}
and one can see that for $t=t_r/4$ the single particle phases phases become independent of $n$
\begin{align}
    \phi_n^e(t_r/4)&= \frac{\pi}{2} \pmod {2\pi}\\ 
    \phi_n^o(t_r/4)&= 0 \pmod {2\pi},
\end{align}
and in particular are purely real or imaginary. This means that the orthonormality that was present at $t=0$ between the states is recovered at integer multiples of $t_r/4$ and the overlap matrix $\mathcal{O}_{kl}(t)$ contains only diagonal elements. For a quench $V^{(1)}\rightarrow V^{(2)}$ (or every quench where the difference in the order of the supersymmetric partner potentials is odd) the diagonal elements describe a transition from a state with an even quantum number to one with an odd quantum number (in the original basis) and the phases of the overlap matrix elements in Eq.~\eqref{eq:overlap_matrix} can be calculated to be
\begin{equation}
    \label{eq:unlike_phase}
    \Delta\phi = \phi^\text{e,o}\left( \frac{t_r}{4}\right) - \phi^\text{o,e}\left(\frac{t_r}{4}\right) = \pm\frac{\pi}{2} \pmod {2\pi}
\end{equation}
for all possible combinations of overlaps.
 
For the quench $V^{(1)}\rightarrow V^{(3)}$ (or every quench where the difference in the order of the supersymmetric partner potentials is even), the corresponding phase difference is the difference between like parity states, which is simply zero or a multiple of $2\pi$. Since in both cases the determinant is just the absolute value of the product of the diagonal elements, the revivals in the survival probability in Eq.~\eqref{eq:survival_probability_T0} occur.

From these insights above one can classify the revivals of the survival probability into two different groups \cite{Revivals:infinite_explained:2000}. For times where the phases of the overlaps do not align at a multiple of $2\pi$ the wavefunction density overlaps with the density at $t =0$, however the imaginary and real parts of the wavefunction do not align. Such revivals are called quasi-revivals. Conversely, at integer multiples of $t_r$ the overlap phases align at multiples of $2\pi$ and are called true revivals.

In between the full revivals the dynamics in each quench is different and one can see that for higher order quenches the survival probability is approaching zero in these regions. It is worth noting that the initial decay of the survival probability is directly related to the average work done on the system \cite{fusco2014}, which can therefore be used to quantify the strength of the quench. For the quenches between supersymmetric Hamiltonians we can calculate the quantity explicitly to be (taking advantage of the fact that we only need to know properties of the initial Hamiltonian)
\begin{align} 
    \langle W^{(\alpha)} \rangle &= -i \frac{d}{dt} \mathcal{O}(t)\big\vert_{t=0}\nonumber\\
    &=\sum_m\left( \mathcal{E}^{(\alpha)}_m-\mathcal{E}^{(1)}_0 \right) \left\vert\langle \Psi_m^{(\alpha)} | \Psi_0^{(1)} \rangle\right\vert^2\nonumber\\
    &=E^{(1)}_1 N(N+1)(\alpha^2 -\alpha).
\end{align}
From this expression one can immediately see that quenches to higher order potentials result in larger amounts of work. Furthermore, the irreversible work can quantify the amount of non-equilibrium excitations induced by the quench and can be calculated to be
\begin{equation}
    \langle W_{\text{irr}}^{(\alpha)}\rangle = \langle W^{(\alpha)}\rangle - \Delta \mathcal{E}^{(\alpha)}_0= E_1 N^2(\alpha -1)^2\,,
\end{equation}
where
\begin{align}
    \Delta \mathcal{E}_0^{(\alpha)}&= \sum_{l=1}^{N}E_{l+\alpha-1}^{(1)}-\sum_{l=1}^{N}E_{l}^{(1)}\nonumber\\
    &=(\alpha-1)(N^2+\alpha N)E^{(1)}_1 
\end{align}
is the energy difference between the ground states of the initial and the final potential.

\begin{figure}[tb]
    \centering
    \includegraphics[width =\linewidth]{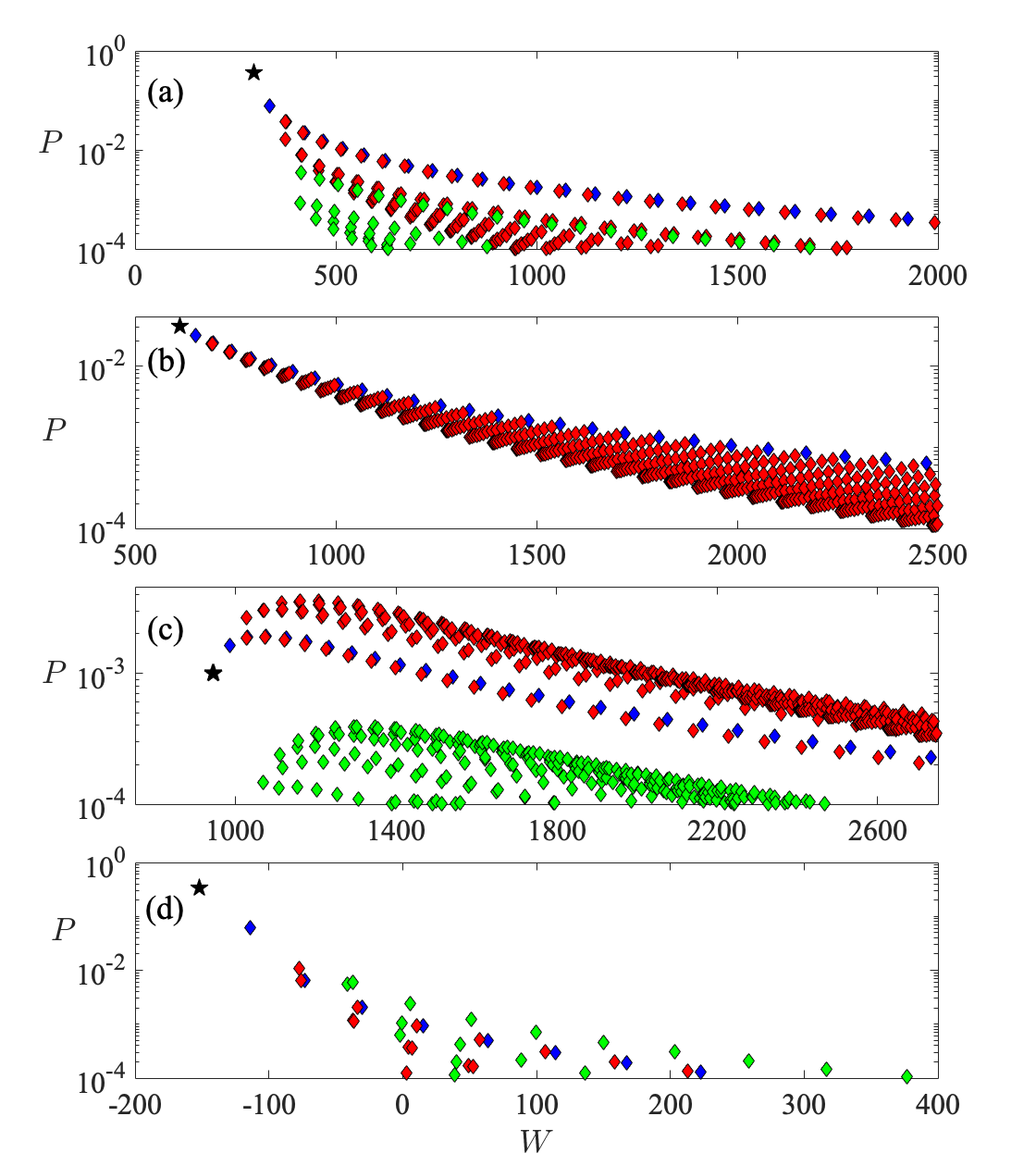}
    \caption{Work probability distribution corresponding to the quenches shown in Fig.~\ref{fig:SurvivalT0}. The transition between ground states of both potentials is marked with a black star, with other transitions colour coded for single particle transitions (blue), two particle transitions (red) and three particle transitions (green).}
    \label{fig:Work_dist}
\end{figure}

Additional insight into the quench dynamics can be obtained by examining the work probability distribution (WPD). This quantity is given by the Fourier transform of the dynamical overlap  
\begin{align}
    P(W)&=\int dt \, e^{-i W t } \mathcal{O}(t)\\
    &=\sum_m\left\vert\langle \Psi_m^{(\alpha)} | \Psi_0^{(1)} \rangle\right\vert^2 \delta\left(W-\left( \mathcal{E}^{(\alpha)}_m-\mathcal{E}^{(1)}_0 \right)  \right)\,,
    \label{eq:WPD}
\end{align}
and allows one to understand how the excitations are spread in the Hilbert space of the final Hamiltonian \cite{Fluc:2011}. We show this quantity in Fig.~\ref{fig:Work_dist}(a-c) for the supersymmetric quenches corresponding to the quench dynamics shown in Fig.~\ref{fig:SurvivalT0}(a-c). The weakest quench is to $V^{(2)}$, where the survival probability is small, but still finite, in between revivals. This indicates that the initial state is not completely destroyed by the quench and, correspondingly, the WPD shows a significant peak at $\Delta\mathcal{E}_0^{(2)}$ (black star in Fig.~\ref{fig:Work_dist}(a)). This is the lowest possible excitation, which corresponds to the transition probability between the respective groundstates $|\Psi_0^{(1)}\rangle$ and $|\Psi_0^{(2)}\rangle$. One can also distinguish the higher energy excitations through the number of particles that are excited from the initial Fermi sea: a single particle excitation (blue diamonds) corresponds to the lowest $N-1$ states being occupied and one particle occupying a state above $N$; a two particle excitation (red diamonds) leads to the lowest $N-2$ states being occupied and two particles occupying states above $N-1$, and higher order excitations are defined  similarly. 

One can see from Fig.~\ref{fig:Work_dist}(a) that single particle excitations possess the largest probability and exist in a single band. For  excitations consisting of more particles there is a larger set of different two particle combinations, however the majority of these can be seen to possess a lower probability, similarly for the three particle excitations. For the stronger quenches to $V^{(3)}$ and $V^{(4)}$, the WPD broadens and the probability to occupy the groundstate $|\langle \Psi_0^{(\alpha)}|\Psi_0^{(1)}\rangle|^2$ decreases (black star in panels \ref{fig:Work_dist}(b) and (c)), which corresponds to the survival probability in Fig.~\ref{fig:SurvivalT0} essentially vanishing between revivals.  In fact, the WPD for these quenches shows that two particle excitations dominate the dynamics and single particle excitations become less important in the quench to $V^{(4)}$. Nevertheless, for each supersymmetric quench an underlying structure can be observed in the WPD which is related to the appearance of the perfect revivals. While for weaker quenches the majority of the excitations are happening just above the Fermi surface, for stronger quenches this is no longer true.

Finally, it is interesting to compare the quench dynamics observed through $F(t)$ to the well-known Talbot effect. In this an initial state is released into a box potential and in the resulting dynamics repeated images of the initial state are formed at regular and well defined times \cite{Bluhm:WF_Revival:1996,Revivals:1997,Revivals:infinite_explained:2000,infinite_Talbot:1997,SpaceT_infinte:1997}. In Fig.~\ref{fig:SurvivalT0}(d) we show the survival probability for a Fermi gas after a sudden expansion of the infinite well potential. The initial state is the ground state before the expansion and one can see that also here the full revivals occur at integer multiples of $t=t_r/4$ owing to the quadratic energy spectrum of the final potential. Note that, for a fair comparison, we have chosen the change in the box-width such that the survival probability in between the revivals is visually close to the one from the supersymmetric quench to $V^{(2)}$. Nevertheless, the WPD shown in Fig.~\ref{fig:Work_dist}(d) can be seen to possess quite a different structure, indicating that the supersymmetric quench is intrinsically different to the Talbot effect. This will become even more clear when looking at quenches at finite temperatures.

\section{Quench at Finite Temperature}

\begin{figure}[tb]
    \centering
    \includegraphics[width =0.9\linewidth]{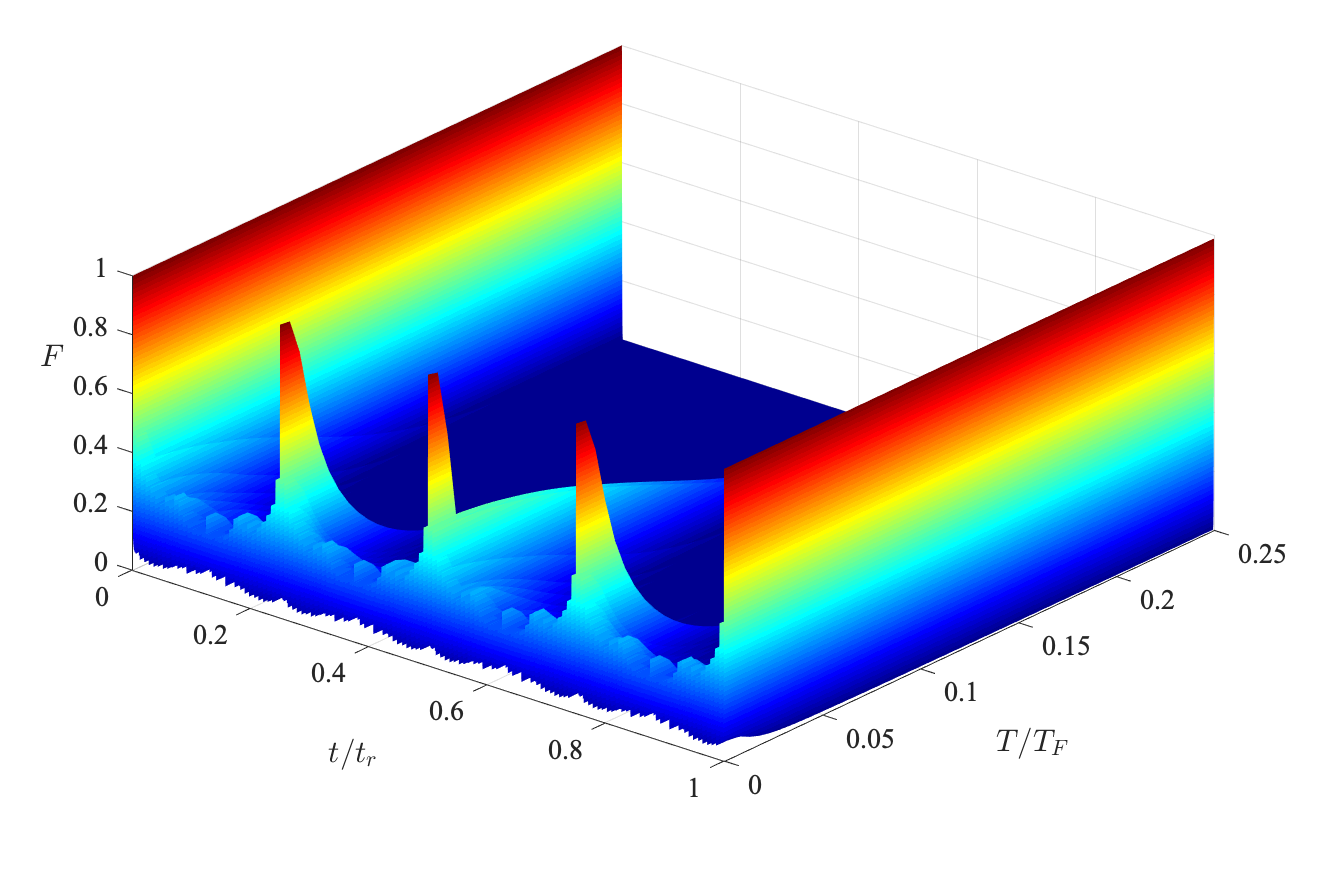}
    \caption{Survival probability $F(t)$ as a function of temperature and time for a quench $V^{(1)}\rightarrow V^{(2)}$.}
    \label{fig:surface_temp_linear}
\end{figure}

While the revivals in the survival probability at zero temperature can be connected to the existence of a Fermi-sea (all occupation probabilities in Eq.~\eqref{eq:survival_probability} are either zero or one) which restricts the dynamics of all lower lying particles \cite{Busch:PB:1998,PBlock:2001}, the behaviour at finite temperatures can be more complicated: with increasing temperature the Fermi-edge becomes less sharp and the occupation probabilities for states below the Fermi edge decreases below one. This expands the number of possible transitions after the quench and therefore leads to a stronger decay of the initial state \cite{Demler:ORTHO:2012,MOSSY:Thermo:2020}. 

\begin{figure}[tb]
    \centering
    \includegraphics[width =0.9\linewidth]{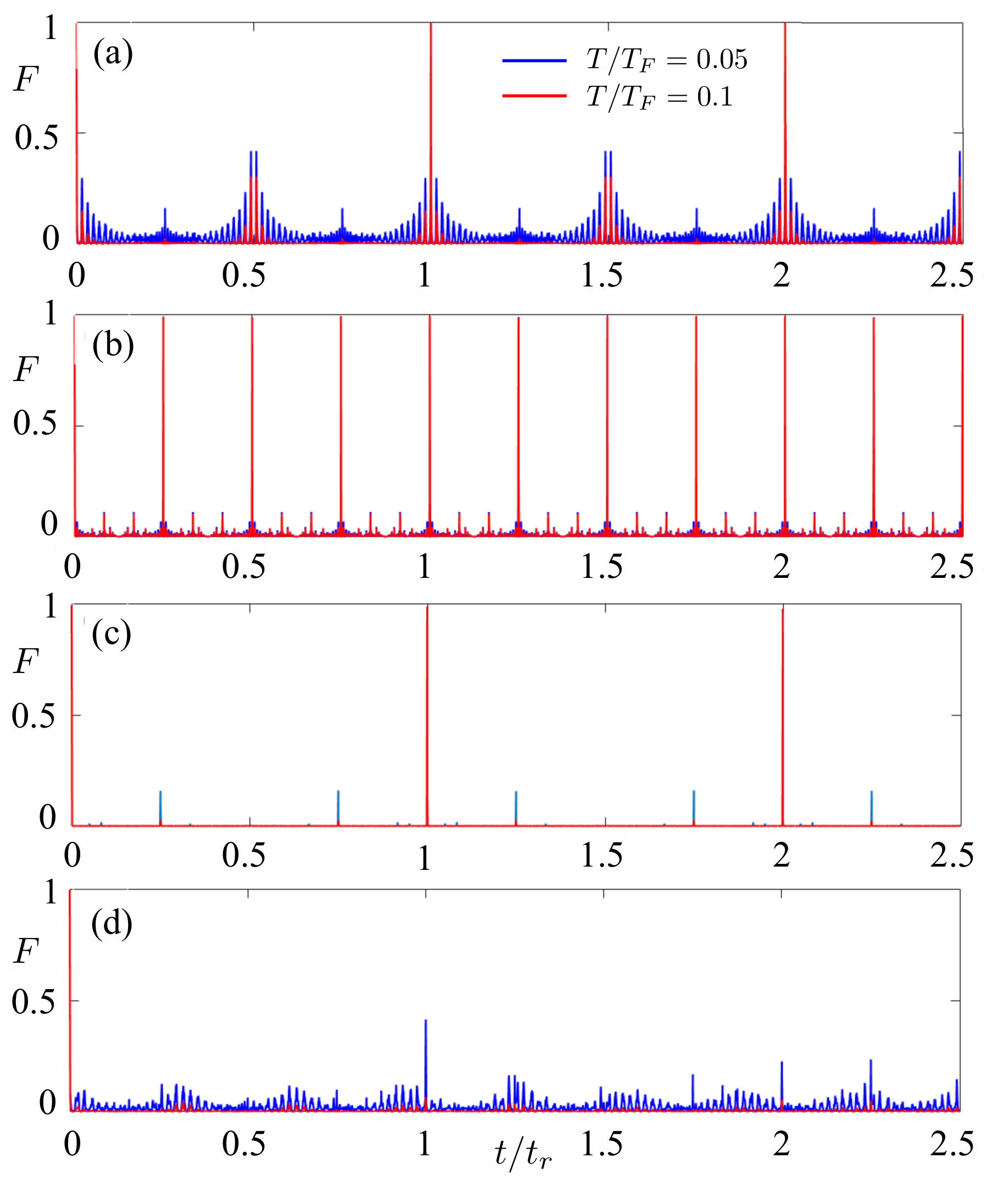}
    \caption{Survival probability for finite temperatures of $T/T_F = 0.05$ (blue) and $T/T_F = 0.1$ (red) for the same quenches as presented in Fig.~\ref{fig:SurvivalT0}.}
    \label{fig:SurvPropT}
\end{figure}

In Fig.~\ref{fig:surface_temp_linear} we show the survival probability for the quench  $V^{(1)}\rightarrow V^{(2)}$ as a function of time and temperature in units of the Fermi temperature $T_F=N^2E_1/k_B$. One can immediately see that the revivals at integer multiples of $t_r$ are not affected by temperature, but all other ones present at zero temperature decay as the temperature increases. For longer times this is also shown in Fig.~\ref{fig:SurvPropT}(a) and a similar behaviour can be observed for a quench $V^{(1)}\rightarrow V^{(4)}$ in Fig.~\ref{fig:SurvPropT}(c). However, for the quench  $V^{(1)}\rightarrow V^{(3)}$ all the revivals are unaffected by temperature. This phenomenon is also observed for quenches between potentials in a hierarchy from $V^{(\alpha)} \rightarrow V^{(\alpha + 2n)}$.
 
\begin{figure}[tb]
    \centering
    \includegraphics[width =\linewidth]{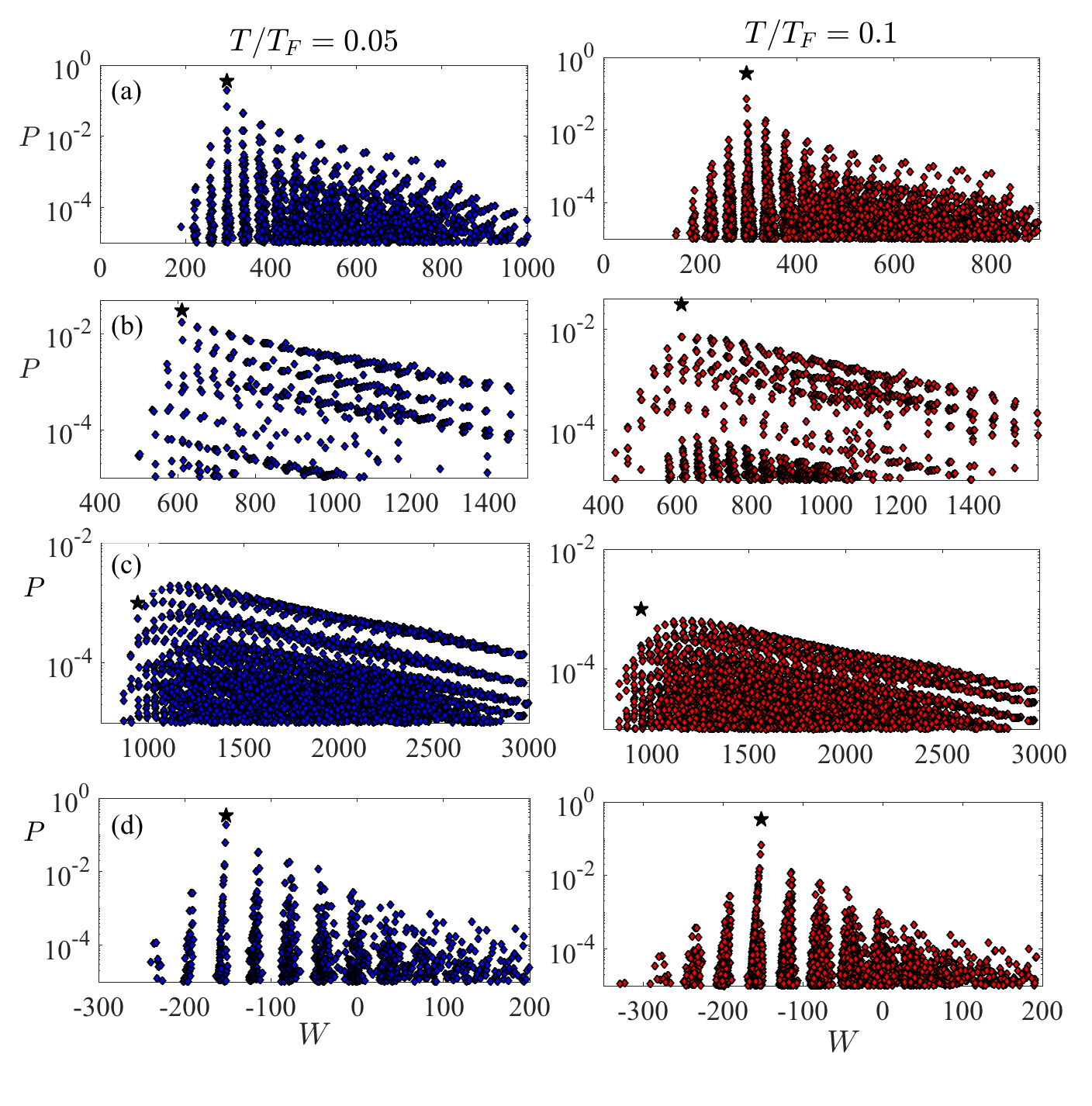}
    \caption{Survival probability and work probability plotted for finite temperatures of $T/T_F = 0.05$ and $T/T_F = 0.1$ for the same quenches as presented in Fig.~\ref{fig:SurvivalT0}. The black star is the same as in Fig.~\ref{fig:Work_dist} (for $T=0$) and included for easier comparison.}
    \label{fig:Log_scale_finite_work}
\end{figure}

To calculate the WPD for finite temperatures we include the thermal ensemble of the initial state $p^{(1)}_{n} = \frac{1}{Z_0}e^{\beta(\mathcal{E}^{(1)}_{n}-N\mu)}$, which in analogy with Eq.~\eqref{eq:WPD} gives
\begin{equation}
    \label{work_dist}
    P(W) = \sum_{m,n} p^{(1)}_{n} \left\vert\langle \Psi_m^{(\alpha)} | \Psi_n^{(1)} \rangle\right\vert^2 \delta\left(W-\left( \mathcal{E}^{(\alpha)}_m-\mathcal{E}^{(1)}_n \right)  \right)\,.
\end{equation}
In Fig.~\ref{fig:Log_scale_finite_work} one can see that, when compared to the zero temperature case, the distributions for the quenches to $V^{(2)}$ and $V^{(4)}$ are much denser and less structured, while for the quench to the potential $V^{(3)}$ the distribution remains comparatively sparse and ordered.

\begin{figure}[tb]
    \includegraphics[width =\linewidth]{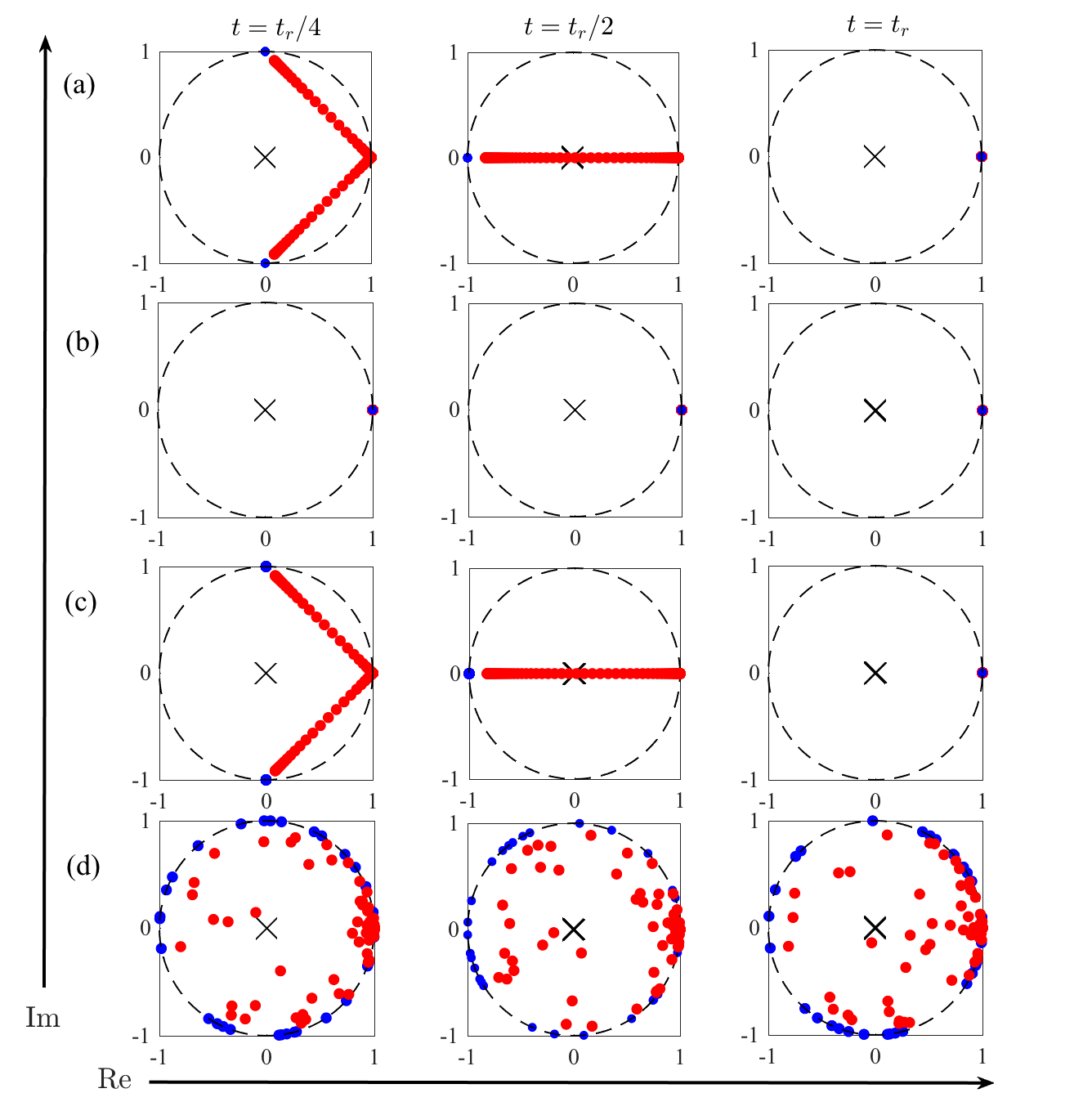}
    \caption{Diagonal elements of the  overlaps for each quench at $T/T_F = 0$ (blue dots) and finite temperature of $T/T_F = 0.5$ (red dots). (a-c) show supersymmetric quenches where target potentials are partner potentials $V^{(2)}$, $V^{(3)}$ and $V^{(4)}$ respectively and (d) shows the expanding box. The off-diagonal elements are zero at these times, which is indicted by the black crosses in the center of each panel.}
    \label{fig:phase_factor_plot}
\end{figure}

At finite temperatures the difference between the supersymmetric quenches and the Talbot effect in a quenched expanding box becomes especially clear.
One can see from Fig.~\ref{fig:SurvPropT}(d) that, despite the similarity in the eigenspectra of these two class of quenches, all revivals completely disappear for the box expansion.  This difference is not obvious from the WPD (see Fig.~\ref{fig:Log_scale_finite_work}(d)) which shows similar broadening of the probability distribution and an increase in the number of excitations. Instead, the presence and absence of constructive interference of the single particle states at the revival times can be best seen by looking at the elements of the determinant in Eq.~\eqref{eq:survival_probability} at integer multiples of $t_r/4$. For full revivals of the initial state all off-diagonal elements should vanish and the remaining diagonal elements should sit on the unit circle. While this is true at zero temperature as shown in Fig.~\ref{fig:phase_factor_plot} (blue dots show the diagonal elements, whereas the cross indicates the off-diagonal ones), at finite temperatures (red dots) this is no longer the case for the supersymmetric quenches $V_1\rightarrow V_2$ and $V_1\rightarrow V_4$ at $t_r/4$ and $t_r/2$, as well as the quench of the box size. This is a consequence of the finite temperature Fermi--Dirac distribution in Eq.~\eqref{eq:survival_probability} which means that the diagonal elements of the determinant are no longer just $\pm 1$ or $\pm i$, and therefore the survival probability is smaller than one.

However, for the supersymmetric quench $V_1\rightarrow V_3$ at any multiple of $t_r/4$ and for all supersymmetric quenches at $t_r$, the elements of the determinant are lying on the unit circle with a well defined phase of $2\pi$. This corresponds to the revivals of the exact single particle wavefunctions at these times, and these real revivals are independent of temperature. Therefore for the supersymmetric quenches only the quasi-revivals are affected by finite temperatures. This also  sheds light on why all revivals vanish for the Talbot quench, despite the energy spectrum having a quadratic form before and after the quench. From Fig.~\ref{fig:phase_factor_plot}(d) one can see that at $T=0$ the diagonal elements are distributed around the unit circle (blue dots) at any integer multiple of $t_r/4$, indicating that all revivals are just quasi-revivals. This is consistent with the observation that at finite temperatures (red dots) the elements of the determinant no longer have a magnitude of one, and therefore the revivals vanish. This highlights the resilience of the supersymmetric systems to thermal effects.

\section{Conclusion}
In this work we have investigated the many-body out-of-equilibrium dynamics for a gas of spin-polarised fermions following a sudden quench in the trapping potential. In particular we have focused on quenches between an infinite box potential to a hierarchy of supersymmetric potentials which are connected through a supersymmetric algebra. What is unique about this scenario is the similarity of the eigenspectra between these different potentials, which results in interesting and accessible non-equilibrium dynamics. Indeed we have shown that quenches between these supersymmetric potentials at zero temperature produce periodic revivals in the survival probability and compared this to the revivals present in Talbot oscillations in the infinite well. Quenches to higher order potentials lead to a stronger decay of the survival probability in between these revivals, which can be understood by quantifying the post-quench excitations through the average work done. We have shown that, due to the existence of the intertwining properties between supersymmetric systems, the average and the irreversible work at zero temperature can be exactly calculated. For finite temperatures we have observed that the quasi-revivals decay in the many-body dynamics, whereas the true revivals remain and are temperature independent. These different revivals can be identified by analyzing the elements of the overlap matrix at zero and finite temperatures, showing that the presence of perfect revivals at finite temperature is strongly dependent on the order of the potential quench.   

Comparing these supersymmetric quenches to the well known Talbot oscillations, where a quench corresponds to changing the width of an infinite box, it is clear that supersymmetric quenches of such potentials are fundamentally different, possess interesting many-body dynamics, and are highly analytically accessible. Supersymmetric quenches are therefore an interesting addition to better understand non-equilibrium dynamics of quantum many-body systems. Finally, while our work has focused on quenches from the infinite box potential, a similar analysis can be readily extended to quenches between other well known superpotentials \cite{COOPER:SUSYBIBLE:1997}. However, as these possess a different structure of the energy spectrum, many-body revivals might not be present and one would have to look for other distinctive measures. This poses interesting questions about the role of supersymmetry in different correlation functions and related concepts in information scrambling that we leave for future work. 

\acknowledgements
This work was supported by the Okinawa Institute of Science and Technology Graduate University and utilized the computing resources of the Scientific Computing and Data Analysis section of the Research Support Division at OIST. TF acknowledges support under JSPS KAKENHI-21K13856.

\bibliography{biblo.bib}

\begin{thebibliography}{59}%
\makeatletter
\providecommand \@ifxundefined [1]{%
 \@ifx{#1\undefined}
}%
\providecommand \@ifnum [1]{%
 \ifnum #1\expandafter \@firstoftwo
 \else \expandafter \@secondoftwo
 \fi
}%
\providecommand \@ifx [1]{%
 \ifx #1\expandafter \@firstoftwo
 \else \expandafter \@secondoftwo
 \fi
}%
\providecommand \natexlab [1]{#1}%
\providecommand \enquote  [1]{``#1''}%
\providecommand \bibnamefont  [1]{#1}%
\providecommand \bibfnamefont [1]{#1}%
\providecommand \citenamefont [1]{#1}%
\providecommand \href@noop [0]{\@secondoftwo}%
\providecommand \href [0]{\begingroup \@sanitize@url \@href}%
\providecommand \@href[1]{\@@startlink{#1}\@@href}%
\providecommand \@@href[1]{\endgroup#1\@@endlink}%
\providecommand \@sanitize@url [0]{\catcode `\\12\catcode `\$12\catcode
  `\&12\catcode `\#12\catcode `\^12\catcode `\_12\catcode `\%12\relax}%
\providecommand \@@startlink[1]{}%
\providecommand \@@endlink[0]{}%
\providecommand \url  [0]{\begingroup\@sanitize@url \@url }%
\providecommand \@url [1]{\endgroup\@href {#1}{\urlprefix }}%
\providecommand \urlprefix  [0]{URL }%
\providecommand \Eprint [0]{\href }%
\providecommand \doibase [0]{http://dx.doi.org/}%
\providecommand \selectlanguage [0]{\@gobble}%
\providecommand \bibinfo  [0]{\@secondoftwo}%
\providecommand \bibfield  [0]{\@secondoftwo}%
\providecommand \translation [1]{[#1]}%
\providecommand \BibitemOpen [0]{}%
\providecommand \bibitemStop [0]{}%
\providecommand \bibitemNoStop [0]{.\EOS\space}%
\providecommand \EOS [0]{\spacefactor3000\relax}%
\providecommand \BibitemShut  [1]{\csname bibitem#1\endcsname}%
\let\auto@bib@innerbib\@empty
\bibitem [{\citenamefont {Mistakidis}\ \emph {et~al.}(2022)\citenamefont
  {Mistakidis}, \citenamefont {Volosniev}, \citenamefont {Barfknecht},
  \citenamefont {Fogarty}, \citenamefont {Busch}, \citenamefont {Foerster},
  \citenamefont {Schmelcher},\ and\ \citenamefont {Zinner}}]{Mistakidis:22}%
  \BibitemOpen
  \bibfield  {author} {\bibinfo {author} {\bibfnamefont {S.~I.}\ \bibnamefont
  {Mistakidis}}, \bibinfo {author} {\bibfnamefont {A.~G.}\ \bibnamefont
  {Volosniev}}, \bibinfo {author} {\bibfnamefont {R.~E.}\ \bibnamefont
  {Barfknecht}}, \bibinfo {author} {\bibfnamefont {T.}~\bibnamefont {Fogarty}},
  \bibinfo {author} {\bibfnamefont {Th.}\ \bibnamefont {Busch}}, \bibinfo
  {author} {\bibfnamefont {A.}~\bibnamefont {Foerster}}, \bibinfo {author}
  {\bibfnamefont {P.}~\bibnamefont {Schmelcher}}, \ and\ \bibinfo {author}
  {\bibfnamefont {N.~T.}\ \bibnamefont {Zinner}},\ }\href@noop {} {\enquote
  {\bibinfo {title} {{Cold atoms in low dimensions -- a laboratory for quantum
  dynamics}},}\ } (\bibinfo {year} {2022}),\ \Eprint
  {http://arxiv.org/abs/2202.11071} {arXiv:2202.11071 [cond-mat.quant-gas]}
  \BibitemShut {NoStop}%
\bibitem [{\citenamefont {Eisert}\ \emph {et~al.}(2015)\citenamefont {Eisert},
  \citenamefont {Friesdorf},\ and\ \citenamefont {Gogolin}}]{Eisert:15}%
  \BibitemOpen
  \bibfield  {author} {\bibinfo {author} {\bibfnamefont {J.}~\bibnamefont
  {Eisert}}, \bibinfo {author} {\bibfnamefont {M.}~\bibnamefont {Friesdorf}}, \
  and\ \bibinfo {author} {\bibfnamefont {C.}~\bibnamefont {Gogolin}},\
  }\bibfield  {title} {\enquote {\bibinfo {title} {Quantum many-body systems
  out of equilibrium},}\ }\href {\doibase 10.1038/nphys3215} {\bibfield
  {journal} {\bibinfo  {journal} {Nature Physics}\ }\textbf {\bibinfo {volume}
  {11}},\ \bibinfo {pages} {124--130} (\bibinfo {year} {2015})}\BibitemShut
  {NoStop}%
\bibitem [{\citenamefont {Kinoshita}\ \emph {et~al.}(2006)\citenamefont
  {Kinoshita}, \citenamefont {Wenger},\ and\ \citenamefont
  {Weiss}}]{Kinoshita:06}%
  \BibitemOpen
  \bibfield  {author} {\bibinfo {author} {\bibfnamefont {Toshiya}\ \bibnamefont
  {Kinoshita}}, \bibinfo {author} {\bibfnamefont {Trevor}\ \bibnamefont
  {Wenger}}, \ and\ \bibinfo {author} {\bibfnamefont {David~S.}\ \bibnamefont
  {Weiss}},\ }\bibfield  {title} {\enquote {\bibinfo {title} {{A quantum
  Newton's cradle}},}\ }\href {\doibase 10.1038/nature04693} {\bibfield
  {journal} {\bibinfo  {journal} {Nature}\ }\textbf {\bibinfo {volume} {440}},\
  \bibinfo {pages} {900--903} (\bibinfo {year} {2006})}\BibitemShut {NoStop}%
\bibitem [{\citenamefont {del Campo}\ and\ \citenamefont
  {Muga}(2006)}]{Campo_2006}%
  \BibitemOpen
  \bibfield  {author} {\bibinfo {author} {\bibfnamefont {A.}~\bibnamefont {del
  Campo}}\ and\ \bibinfo {author} {\bibfnamefont {J.~G}\ \bibnamefont {Muga}},\
  }\bibfield  {title} {\enquote {\bibinfo {title} {{Dynamics of a
  Tonks-Girardeau gas released from a hard-wall trap}},}\ }\href {\doibase
  10.1209/epl/i2006-10061-5} {\bibfield  {journal} {\bibinfo  {journal}
  {Europhysics Letters ({EPL})}\ }\textbf {\bibinfo {volume} {74}},\ \bibinfo
  {pages} {965--971} (\bibinfo {year} {2006})}\BibitemShut {NoStop}%
\bibitem [{\citenamefont {Collura}\ \emph {et~al.}(2013)\citenamefont
  {Collura}, \citenamefont {Sotiriadis},\ and\ \citenamefont
  {Calabrese}}]{Collura13}%
  \BibitemOpen
  \bibfield  {author} {\bibinfo {author} {\bibfnamefont {Mario}\ \bibnamefont
  {Collura}}, \bibinfo {author} {\bibfnamefont {Spyros}\ \bibnamefont
  {Sotiriadis}}, \ and\ \bibinfo {author} {\bibfnamefont {Pasquale}\
  \bibnamefont {Calabrese}},\ }\bibfield  {title} {\enquote {\bibinfo {title}
  {{Equilibration of a Tonks-Girardeau Gas Following a Trap Release}},}\ }\href
  {\doibase 10.1103/PhysRevLett.110.245301} {\bibfield  {journal} {\bibinfo
  {journal} {Phys. Rev. Lett.}\ }\textbf {\bibinfo {volume} {110}},\ \bibinfo
  {pages} {245301} (\bibinfo {year} {2013})}\BibitemShut {NoStop}%
\bibitem [{\citenamefont {Nagao}\ \emph {et~al.}(2019)\citenamefont {Nagao},
  \citenamefont {Kunimi}, \citenamefont {Takasu}, \citenamefont {Takahashi},\
  and\ \citenamefont {Danshita}}]{Kazuma:19}%
  \BibitemOpen
  \bibfield  {author} {\bibinfo {author} {\bibfnamefont {Kazuma}\ \bibnamefont
  {Nagao}}, \bibinfo {author} {\bibfnamefont {Masaya}\ \bibnamefont {Kunimi}},
  \bibinfo {author} {\bibfnamefont {Yosuke}\ \bibnamefont {Takasu}}, \bibinfo
  {author} {\bibfnamefont {Yoshiro}\ \bibnamefont {Takahashi}}, \ and\ \bibinfo
  {author} {\bibfnamefont {Ippei}\ \bibnamefont {Danshita}},\ }\bibfield
  {title} {\enquote {\bibinfo {title} {{Semiclassical quench dynamics of Bose
  gases in optical lattices}},}\ }\href {\doibase 10.1103/PhysRevA.99.023622}
  {\bibfield  {journal} {\bibinfo  {journal} {Phys. Rev. A}\ }\textbf {\bibinfo
  {volume} {99}},\ \bibinfo {pages} {023622} (\bibinfo {year}
  {2019})}\BibitemShut {NoStop}%
\bibitem [{\citenamefont {Mistakidis}\ \emph {et~al.}(2014)\citenamefont
  {Mistakidis}, \citenamefont {Cao},\ and\ \citenamefont
  {Schmelcher}}]{Mistakidis_2014}%
  \BibitemOpen
  \bibfield  {author} {\bibinfo {author} {\bibfnamefont {S.~I.}\ \bibnamefont
  {Mistakidis}}, \bibinfo {author} {\bibfnamefont {L.}~\bibnamefont {Cao}}, \
  and\ \bibinfo {author} {\bibfnamefont {P.}~\bibnamefont {Schmelcher}},\
  }\bibfield  {title} {\enquote {\bibinfo {title} {Interaction quench induced
  multimode dynamics of finite atomic ensembles},}\ }\href {\doibase
  10.1088/0953-4075/47/22/225303} {\bibfield  {journal} {\bibinfo  {journal}
  {Journal of Physics B: Atomic, Molecular and Optical Physics}\ }\textbf
  {\bibinfo {volume} {47}},\ \bibinfo {pages} {225303} (\bibinfo {year}
  {2014})}\BibitemShut {NoStop}%
\bibitem [{\citenamefont {Garc{\'{\i}}a-March}\ \emph
  {et~al.}(2016)\citenamefont {Garc{\'{\i}}a-March}, \citenamefont {Fogarty},
  \citenamefont {Campbell}, \citenamefont {Busch},\ and\ \citenamefont
  {Paternostro}}]{GarciaMarch:16}%
  \BibitemOpen
  \bibfield  {author} {\bibinfo {author} {\bibfnamefont {Miguel~{\'{A}}ngel}\
  \bibnamefont {Garc{\'{\i}}a-March}}, \bibinfo {author} {\bibfnamefont
  {Thom{\'{a}}s}\ \bibnamefont {Fogarty}}, \bibinfo {author} {\bibfnamefont
  {Steve}\ \bibnamefont {Campbell}}, \bibinfo {author} {\bibfnamefont {Thomas}\
  \bibnamefont {Busch}}, \ and\ \bibinfo {author} {\bibfnamefont {Mauro}\
  \bibnamefont {Paternostro}},\ }\bibfield  {title} {\enquote {\bibinfo {title}
  {Non-equilibrium thermodynamics of harmonically trapped bosons},}\ }\href
  {\doibase 10.1088/1367-2630/18/10/103035} {\bibfield  {journal} {\bibinfo
  {journal} {New Journal of Physics}\ }\textbf {\bibinfo {volume} {18}},\
  \bibinfo {pages} {103035} (\bibinfo {year} {2016})}\BibitemShut {NoStop}%
\bibitem [{\citenamefont {Knap}\ \emph
  {et~al.}(2012{\natexlab{a}})\citenamefont {Knap}, \citenamefont {Shashi},
  \citenamefont {Nishida}, \citenamefont {Imambekov}, \citenamefont {Abanin},\
  and\ \citenamefont {Demler}}]{Knap:12}%
  \BibitemOpen
  \bibfield  {author} {\bibinfo {author} {\bibfnamefont {Michael}\ \bibnamefont
  {Knap}}, \bibinfo {author} {\bibfnamefont {Aditya}\ \bibnamefont {Shashi}},
  \bibinfo {author} {\bibfnamefont {Yusuke}\ \bibnamefont {Nishida}}, \bibinfo
  {author} {\bibfnamefont {Adilet}\ \bibnamefont {Imambekov}}, \bibinfo
  {author} {\bibfnamefont {Dmitry~A.}\ \bibnamefont {Abanin}}, \ and\ \bibinfo
  {author} {\bibfnamefont {Eugene}\ \bibnamefont {Demler}},\ }\bibfield
  {title} {\enquote {\bibinfo {title} {{Time-Dependent Impurity in Ultracold
  Fermions: Orthogonality Catastrophe and Beyond}},}\ }\href {\doibase
  10.1103/PhysRevX.2.041020} {\bibfield  {journal} {\bibinfo  {journal} {Phys.
  Rev. X}\ }\textbf {\bibinfo {volume} {2}},\ \bibinfo {pages} {041020}
  (\bibinfo {year} {2012}{\natexlab{a}})}\BibitemShut {NoStop}%
\bibitem [{\citenamefont {Campbell}\ \emph {et~al.}(2014)\citenamefont
  {Campbell}, \citenamefont {Garc\'{\i}a-March}, \citenamefont {Fogarty},\ and\
  \citenamefont {Busch}}]{Campbell:14}%
  \BibitemOpen
  \bibfield  {author} {\bibinfo {author} {\bibfnamefont {Steve}\ \bibnamefont
  {Campbell}}, \bibinfo {author} {\bibfnamefont {Miguel~\'Angel}\ \bibnamefont
  {Garc\'{\i}a-March}}, \bibinfo {author} {\bibfnamefont {Thom\'as}\
  \bibnamefont {Fogarty}}, \ and\ \bibinfo {author} {\bibfnamefont {Thomas}\
  \bibnamefont {Busch}},\ }\bibfield  {title} {\enquote {\bibinfo {title}
  {Quenching small quantum gases: Genesis of the orthogonality catastrophe},}\
  }\href {\doibase 10.1103/PhysRevA.90.013617} {\bibfield  {journal} {\bibinfo
  {journal} {Phys. Rev. A}\ }\textbf {\bibinfo {volume} {90}},\ \bibinfo
  {pages} {013617} (\bibinfo {year} {2014})}\BibitemShut {NoStop}%
\bibitem [{\citenamefont {Keller}\ and\ \citenamefont
  {Fogarty}(2016)}]{Tim:Quench:2016}%
  \BibitemOpen
  \bibfield  {author} {\bibinfo {author} {\bibfnamefont {Tim}\ \bibnamefont
  {Keller}}\ and\ \bibinfo {author} {\bibfnamefont {Thom\'as}\ \bibnamefont
  {Fogarty}},\ }\bibfield  {title} {\enquote {\bibinfo {title} {Probing the
  out-of-equilibrium dynamics of two interacting atoms},}\ }\href {\doibase
  10.1103/PhysRevA.94.063620} {\bibfield  {journal} {\bibinfo  {journal} {Phys.
  Rev. A}\ }\textbf {\bibinfo {volume} {94}},\ \bibinfo {pages} {063620}
  (\bibinfo {year} {2016})}\BibitemShut {NoStop}%
\bibitem [{\citenamefont {Fogarty}\ \emph {et~al.}(2020)\citenamefont
  {Fogarty}, \citenamefont {Deffner}, \citenamefont {Busch},\ and\
  \citenamefont {Campbell}}]{MOSSY:ORTHO:2020}%
  \BibitemOpen
  \bibfield  {author} {\bibinfo {author} {\bibfnamefont {Thom\'as}\
  \bibnamefont {Fogarty}}, \bibinfo {author} {\bibfnamefont {Sebastian}\
  \bibnamefont {Deffner}}, \bibinfo {author} {\bibfnamefont {Thomas}\
  \bibnamefont {Busch}}, \ and\ \bibinfo {author} {\bibfnamefont {Steve}\
  \bibnamefont {Campbell}},\ }\bibfield  {title} {\enquote {\bibinfo {title}
  {{Orthogonality Catastrophe as a Consequence of the Quantum Speed Limit}},}\
  }\href {\doibase 10.1103/PhysRevLett.124.110601} {\bibfield  {journal}
  {\bibinfo  {journal} {Phys. Rev. Lett.}\ }\textbf {\bibinfo {volume} {124}},\
  \bibinfo {pages} {110601} (\bibinfo {year} {2020})}\BibitemShut {NoStop}%
\bibitem [{\citenamefont {Fusco}\ \emph {et~al.}(2014)\citenamefont {Fusco},
  \citenamefont {Pigeon}, \citenamefont {Apollaro}, \citenamefont {Xuereb},
  \citenamefont {Mazzola}, \citenamefont {Campisi}, \citenamefont {Ferraro},
  \citenamefont {Paternostro},\ and\ \citenamefont {De~Chiara}}]{fusco2014}%
  \BibitemOpen
  \bibfield  {author} {\bibinfo {author} {\bibfnamefont {L.}~\bibnamefont
  {Fusco}}, \bibinfo {author} {\bibfnamefont {S.}~\bibnamefont {Pigeon}},
  \bibinfo {author} {\bibfnamefont {T.~J.~G.}\ \bibnamefont {Apollaro}},
  \bibinfo {author} {\bibfnamefont {A.}~\bibnamefont {Xuereb}}, \bibinfo
  {author} {\bibfnamefont {L.}~\bibnamefont {Mazzola}}, \bibinfo {author}
  {\bibfnamefont {M.}~\bibnamefont {Campisi}}, \bibinfo {author} {\bibfnamefont
  {A.}~\bibnamefont {Ferraro}}, \bibinfo {author} {\bibfnamefont
  {M.}~\bibnamefont {Paternostro}}, \ and\ \bibinfo {author} {\bibfnamefont
  {G.}~\bibnamefont {De~Chiara}},\ }\bibfield  {title} {\enquote {\bibinfo
  {title} {{Assessing the Nonequilibrium Thermodynamics in a Quenched Quantum
  Many-Body System via Single Projective Measurements}},}\ }\href {\doibase
  10.1103/PhysRevX.4.031029} {\bibfield  {journal} {\bibinfo  {journal} {Phys.
  Rev. X}\ }\textbf {\bibinfo {volume} {4}},\ \bibinfo {pages} {031029}
  (\bibinfo {year} {2014})}\BibitemShut {NoStop}%
\bibitem [{\citenamefont {Campbell}(2016)}]{Campbell16}%
  \BibitemOpen
  \bibfield  {author} {\bibinfo {author} {\bibfnamefont {Steve}\ \bibnamefont
  {Campbell}},\ }\bibfield  {title} {\enquote {\bibinfo {title} {{Criticality
  revealed through quench dynamics in the Lipkin-Meshkov-Glick model}},}\
  }\href {\doibase 10.1103/PhysRevB.94.184403} {\bibfield  {journal} {\bibinfo
  {journal} {Phys. Rev. B}\ }\textbf {\bibinfo {volume} {94}},\ \bibinfo
  {pages} {184403} (\bibinfo {year} {2016})}\BibitemShut {NoStop}%
\bibitem [{\citenamefont {Chenu}\ \emph {et~al.}(2018)\citenamefont {Chenu},
  \citenamefont {Egusquiza}, \citenamefont {Molina-Vilaplana},\ and\
  \citenamefont {del Campo}}]{Chenu2018}%
  \BibitemOpen
  \bibfield  {author} {\bibinfo {author} {\bibfnamefont {A.}~\bibnamefont
  {Chenu}}, \bibinfo {author} {\bibfnamefont {I.~L.}\ \bibnamefont
  {Egusquiza}}, \bibinfo {author} {\bibfnamefont {J.}~\bibnamefont
  {Molina-Vilaplana}}, \ and\ \bibinfo {author} {\bibfnamefont
  {A.}~\bibnamefont {del Campo}},\ }\bibfield  {title} {\enquote {\bibinfo
  {title} {{Quantum work statistics, Loschmidt echo and information
  scrambling}},}\ }\href {https://doi.org/10.1038/s41598-018-30982-w}
  {\bibfield  {journal} {\bibinfo  {journal} {Scientific Reports}\ }\textbf
  {\bibinfo {volume} {8}} (\bibinfo {year} {2018})}\BibitemShut {NoStop}%
\bibitem [{\citenamefont {Chenu}\ \emph {et~al.}(2019)\citenamefont {Chenu},
  \citenamefont {Molina-Vilaplana},\ and\ \citenamefont {del
  Campo}}]{Chenu2019workstatistics}%
  \BibitemOpen
  \bibfield  {author} {\bibinfo {author} {\bibfnamefont {Aur{\'{e}}lia}\
  \bibnamefont {Chenu}}, \bibinfo {author} {\bibfnamefont {Javier}\
  \bibnamefont {Molina-Vilaplana}}, \ and\ \bibinfo {author} {\bibfnamefont
  {Adolfo}\ \bibnamefont {del Campo}},\ }\bibfield  {title} {\enquote {\bibinfo
  {title} {Work {S}tatistics, {L}oschmidt {E}cho and {I}nformation {S}crambling
  in {C}haotic {Q}uantum {S}ystems},}\ }\href {\doibase
  10.22331/q-2019-03-04-127} {\bibfield  {journal} {\bibinfo  {journal}
  {{Quantum}}\ }\textbf {\bibinfo {volume} {3}},\ \bibinfo {pages} {127}
  (\bibinfo {year} {2019})}\BibitemShut {NoStop}%
\bibitem [{\citenamefont {Mikkelsen}\ \emph {et~al.}(2022)\citenamefont
  {Mikkelsen}, \citenamefont {Fogarty},\ and\ \citenamefont
  {Busch}}]{Mikkelsen22}%
  \BibitemOpen
  \bibfield  {author} {\bibinfo {author} {\bibfnamefont {M.}~\bibnamefont
  {Mikkelsen}}, \bibinfo {author} {\bibfnamefont {T.}~\bibnamefont {Fogarty}},
  \ and\ \bibinfo {author} {\bibfnamefont {Th.}\ \bibnamefont {Busch}},\
  }\bibfield  {title} {\enquote {\bibinfo {title} {{Connecting Scrambling and
  Work Statistics for Short-Range Interactions in the Harmonic Oscillator}},}\
  }\href {\doibase 10.1103/PhysRevLett.128.070605} {\bibfield  {journal}
  {\bibinfo  {journal} {Phys. Rev. Lett.}\ }\textbf {\bibinfo {volume} {128}},\
  \bibinfo {pages} {070605} (\bibinfo {year} {2022})}\BibitemShut {NoStop}%
\bibitem [{\citenamefont {Deffner}\ \emph {et~al.}(2014)\citenamefont
  {Deffner}, \citenamefont {Jarzynski},\ and\ \citenamefont {del
  Campo}}]{Deffner14}%
  \BibitemOpen
  \bibfield  {author} {\bibinfo {author} {\bibfnamefont {Sebastian}\
  \bibnamefont {Deffner}}, \bibinfo {author} {\bibfnamefont {Christopher}\
  \bibnamefont {Jarzynski}}, \ and\ \bibinfo {author} {\bibfnamefont {Adolfo}\
  \bibnamefont {del Campo}},\ }\bibfield  {title} {\enquote {\bibinfo {title}
  {{Classical and Quantum Shortcuts to Adiabaticity for Scale-Invariant
  Driving}},}\ }\href {\doibase 10.1103/PhysRevX.4.021013} {\bibfield
  {journal} {\bibinfo  {journal} {Phys. Rev. X}\ }\textbf {\bibinfo {volume}
  {4}},\ \bibinfo {pages} {021013} (\bibinfo {year} {2014})}\BibitemShut
  {NoStop}%
\bibitem [{\citenamefont {Minguzzi}\ and\ \citenamefont
  {Gangardt}(2005)}]{Minguzzi05}%
  \BibitemOpen
  \bibfield  {author} {\bibinfo {author} {\bibfnamefont {A.}~\bibnamefont
  {Minguzzi}}\ and\ \bibinfo {author} {\bibfnamefont {D.~M.}\ \bibnamefont
  {Gangardt}},\ }\bibfield  {title} {\enquote {\bibinfo {title} {{Exact
  Coherent States of a Harmonically Confined Tonks-Girardeau Gas}},}\ }\href
  {\doibase 10.1103/PhysRevLett.94.240404} {\bibfield  {journal} {\bibinfo
  {journal} {Phys. Rev. Lett.}\ }\textbf {\bibinfo {volume} {94}},\ \bibinfo
  {pages} {240404} (\bibinfo {year} {2005})}\BibitemShut {NoStop}%
\bibitem [{\citenamefont {Rohringer}\ \emph {et~al.}(2015)\citenamefont
  {Rohringer}, \citenamefont {Fischer}, \citenamefont {Steiner}, \citenamefont
  {Mazets}, \citenamefont {Schmiedmayer},\ and\ \citenamefont
  {Trupke}}]{Rohringer2015}%
  \BibitemOpen
  \bibfield  {author} {\bibinfo {author} {\bibfnamefont {W.}~\bibnamefont
  {Rohringer}}, \bibinfo {author} {\bibfnamefont {D.}~\bibnamefont {Fischer}},
  \bibinfo {author} {\bibfnamefont {F.}~\bibnamefont {Steiner}}, \bibinfo
  {author} {\bibfnamefont {I.~E.}\ \bibnamefont {Mazets}}, \bibinfo {author}
  {\bibfnamefont {J.}~\bibnamefont {Schmiedmayer}}, \ and\ \bibinfo {author}
  {\bibfnamefont {M.}~\bibnamefont {Trupke}},\ }\bibfield  {title} {\enquote
  {\bibinfo {title} {{Non-equilibrium scale invariance and shortcuts to
  adiabaticity in a one-dimensional Bose gas}},}\ }\href
  {https://doi.org/10.1038/srep09820} {\bibfield  {journal} {\bibinfo
  {journal} {Scientific Reports}\ }\textbf {\bibinfo {volume} {5}} (\bibinfo
  {year} {2015})}\BibitemShut {NoStop}%
\bibitem [{\citenamefont {Collura}\ \emph {et~al.}(2018)\citenamefont
  {Collura}, \citenamefont {Kormos},\ and\ \citenamefont
  {Calabrese}}]{Collura18}%
  \BibitemOpen
  \bibfield  {author} {\bibinfo {author} {\bibfnamefont {Mario}\ \bibnamefont
  {Collura}}, \bibinfo {author} {\bibfnamefont {M\'arton}\ \bibnamefont
  {Kormos}}, \ and\ \bibinfo {author} {\bibfnamefont {Pasquale}\ \bibnamefont
  {Calabrese}},\ }\bibfield  {title} {\enquote {\bibinfo {title} {{Quantum
  quench in a harmonically trapped one-dimensional Bose gas}},}\ }\href
  {\doibase 10.1103/PhysRevA.97.033609} {\bibfield  {journal} {\bibinfo
  {journal} {Phys. Rev. A}\ }\textbf {\bibinfo {volume} {97}},\ \bibinfo
  {pages} {033609} (\bibinfo {year} {2018})}\BibitemShut {NoStop}%
\bibitem [{\citenamefont {Friesch}\ \emph {et~al.}(2000)\citenamefont
  {Friesch}, \citenamefont {Marzoli},\ and\ \citenamefont
  {Schleich}}]{Friesch_2000}%
  \BibitemOpen
  \bibfield  {author} {\bibinfo {author} {\bibfnamefont {O.~M.}\ \bibnamefont
  {Friesch}}, \bibinfo {author} {\bibfnamefont {I.}~\bibnamefont {Marzoli}}, \
  and\ \bibinfo {author} {\bibfnamefont {W.~P.}\ \bibnamefont {Schleich}},\
  }\bibfield  {title} {\enquote {\bibinfo {title} {{Quantum carpets woven by
  Wigner functions}},}\ }\href {\doibase 10.1088/1367-2630/2/1/004} {\bibfield
  {journal} {\bibinfo  {journal} {New Journal of Physics}\ }\textbf {\bibinfo
  {volume} {2}},\ \bibinfo {pages} {004} (\bibinfo {year} {2000})}\BibitemShut
  {NoStop}%
\bibitem [{\citenamefont {del Campo}\ and\ \citenamefont
  {Boshier}(2012)}]{Campo2012}%
  \BibitemOpen
  \bibfield  {author} {\bibinfo {author} {\bibfnamefont {A.}~\bibnamefont {del
  Campo}}\ and\ \bibinfo {author} {\bibfnamefont {M.~G.}\ \bibnamefont
  {Boshier}},\ }\bibfield  {title} {\enquote {\bibinfo {title} {Shortcuts to
  adiabaticity in a time-dependent box},}\ }\href
  {https://doi.org/10.1038/srep00648} {\bibfield  {journal} {\bibinfo
  {journal} {Scientific Reports}\ }\textbf {\bibinfo {volume} {2}} (\bibinfo
  {year} {2012})}\BibitemShut {NoStop}%
\bibitem [{\citenamefont {Vicari}(2019)}]{Vicari19}%
  \BibitemOpen
  \bibfield  {author} {\bibinfo {author} {\bibfnamefont {Ettore}\ \bibnamefont
  {Vicari}},\ }\bibfield  {title} {\enquote {\bibinfo {title} {{Particle-number
  scaling of the quantum work statistics and Loschmidt echo in Fermi gases with
  time-dependent traps}},}\ }\href {\doibase 10.1103/PhysRevA.99.043603}
  {\bibfield  {journal} {\bibinfo  {journal} {Phys. Rev. A}\ }\textbf {\bibinfo
  {volume} {99}},\ \bibinfo {pages} {043603} (\bibinfo {year}
  {2019})}\BibitemShut {NoStop}%
\bibitem [{\citenamefont {Grochowski}\ \emph {et~al.}(2020)\citenamefont
  {Grochowski}, \citenamefont {Karpiuk}, \citenamefont {Brewczyk},\ and\
  \citenamefont {Rz\k{a}\ifmmode~\dot{z}\else \.{z}\fi{}ewski}}]{Grochowski20}%
  \BibitemOpen
  \bibfield  {author} {\bibinfo {author} {\bibfnamefont {Piotr~T.}\
  \bibnamefont {Grochowski}}, \bibinfo {author} {\bibfnamefont {Tomasz}\
  \bibnamefont {Karpiuk}}, \bibinfo {author} {\bibfnamefont {Miros\l{}aw}\
  \bibnamefont {Brewczyk}}, \ and\ \bibinfo {author} {\bibfnamefont
  {Kazimierz}\ \bibnamefont {Rz\k{a}\ifmmode~\dot{z}\else \.{z}\fi{}ewski}},\
  }\bibfield  {title} {\enquote {\bibinfo {title} {Fermionic quantum carpets:
  From canals and ridges to solitonlike structures},}\ }\href {\doibase
  10.1103/PhysRevResearch.2.013119} {\bibfield  {journal} {\bibinfo  {journal}
  {Phys. Rev. Research}\ }\textbf {\bibinfo {volume} {2}},\ \bibinfo {pages}
  {013119} (\bibinfo {year} {2020})}\BibitemShut {NoStop}%
\bibitem [{\citenamefont {\L{}ebek}\ \emph {et~al.}(2021)\citenamefont
  {\L{}ebek}, \citenamefont {Grochowski},\ and\ \citenamefont
  {Rz\k{a}\ifmmode~\dot{z}\else \.{z}\fi{}ewski}}]{Revivals:StoM:2021}%
  \BibitemOpen
  \bibfield  {author} {\bibinfo {author} {\bibfnamefont {Maciej}\ \bibnamefont
  {\L{}ebek}}, \bibinfo {author} {\bibfnamefont {Piotr~T.}\ \bibnamefont
  {Grochowski}}, \ and\ \bibinfo {author} {\bibfnamefont {Kazimierz}\
  \bibnamefont {Rz\k{a}\ifmmode~\dot{z}\else \.{z}\fi{}ewski}},\ }\bibfield
  {title} {\enquote {\bibinfo {title} {Single- to many-body crossover of a
  quantum carpet},}\ }\href {\doibase 10.1103/PhysRevResearch.3.023009}
  {\bibfield  {journal} {\bibinfo  {journal} {Phys. Rev. Research}\ }\textbf
  {\bibinfo {volume} {3}},\ \bibinfo {pages} {023009} (\bibinfo {year}
  {2021})}\BibitemShut {NoStop}%
\bibitem [{\citenamefont {Surace}\ \emph {et~al.}(2020)\citenamefont {Surace},
  \citenamefont {Giudici},\ and\ \citenamefont
  {Dalmonte}}]{SUSY_ergodicity:2020}%
  \BibitemOpen
  \bibfield  {author} {\bibinfo {author} {\bibfnamefont {Federica~Maria}\
  \bibnamefont {Surace}}, \bibinfo {author} {\bibfnamefont {Giuliano}\
  \bibnamefont {Giudici}}, \ and\ \bibinfo {author} {\bibfnamefont {Marcello}\
  \bibnamefont {Dalmonte}},\ }\bibfield  {title} {\enquote {\bibinfo {title}
  {Weak-ergodicity-breaking via lattice supersymmetry},}\ }\href {\doibase
  10.22331/q-2020-10-07-339} {\bibfield  {journal} {\bibinfo  {journal}
  {{Quantum}}\ }\textbf {\bibinfo {volume} {4}},\ \bibinfo {pages} {339}
  (\bibinfo {year} {2020})}\BibitemShut {NoStop}%
\bibitem [{\citenamefont {Dooley}\ and\ \citenamefont
  {Kells}(2020)}]{Enhanced_scar:2020}%
  \BibitemOpen
  \bibfield  {author} {\bibinfo {author} {\bibfnamefont {Shane}\ \bibnamefont
  {Dooley}}\ and\ \bibinfo {author} {\bibfnamefont {Graham}\ \bibnamefont
  {Kells}},\ }\bibfield  {title} {\enquote {\bibinfo {title} {Enhancing the
  effect of quantum many-body scars on dynamics by minimizing the effective
  dimension},}\ }\href {\doibase 10.1103/PhysRevB.102.195114} {\bibfield
  {journal} {\bibinfo  {journal} {Phys. Rev. B}\ }\textbf {\bibinfo {volume}
  {102}},\ \bibinfo {pages} {195114} (\bibinfo {year} {2020})}\BibitemShut
  {NoStop}%
\bibitem [{\citenamefont {Turner}\ \emph {et~al.}(2018)\citenamefont {Turner},
  \citenamefont {Michailidis}, \citenamefont {Abanin}, \citenamefont {Serbyn},\
  and\ \citenamefont {Papi{\'c}}}]{Erg_breaking:2018}%
  \BibitemOpen
  \bibfield  {author} {\bibinfo {author} {\bibfnamefont {C.~J.}\ \bibnamefont
  {Turner}}, \bibinfo {author} {\bibfnamefont {A.~A.}\ \bibnamefont
  {Michailidis}}, \bibinfo {author} {\bibfnamefont {D.~A.}\ \bibnamefont
  {Abanin}}, \bibinfo {author} {\bibfnamefont {M.}~\bibnamefont {Serbyn}}, \
  and\ \bibinfo {author} {\bibfnamefont {Z.}~\bibnamefont {Papi{\'c}}},\
  }\bibfield  {title} {\enquote {\bibinfo {title} {Weak ergodicity breaking
  from quantum many-body scars},}\ }\href {\doibase 10.1038/s41567-018-0137-5}
  {\bibfield  {journal} {\bibinfo  {journal} {Nature Physics}\ }\textbf
  {\bibinfo {volume} {14}},\ \bibinfo {pages} {745--749} (\bibinfo {year}
  {2018})}\BibitemShut {NoStop}%
\bibitem [{\citenamefont {Cooper}\ \emph {et~al.}(1995)\citenamefont {Cooper},
  \citenamefont {Khare},\ and\ \citenamefont
  {Sukhatme}}]{COOPER:SUSYBIBLE:1997}%
  \BibitemOpen
  \bibfield  {author} {\bibinfo {author} {\bibfnamefont {Fred}\ \bibnamefont
  {Cooper}}, \bibinfo {author} {\bibfnamefont {Avinash}\ \bibnamefont {Khare}},
  \ and\ \bibinfo {author} {\bibfnamefont {Uday}\ \bibnamefont {Sukhatme}},\
  }\bibfield  {title} {\enquote {\bibinfo {title} {Supersymmetry and quantum
  mechanics},}\ }\href {\doibase https://doi.org/10.1016/0370-1573(94)00080-M}
  {\bibfield  {journal} {\bibinfo  {journal} {Physics Reports}\ }\textbf
  {\bibinfo {volume} {251}},\ \bibinfo {pages} {267--385} (\bibinfo {year}
  {1995})}\BibitemShut {NoStop}%
\bibitem [{\citenamefont {Sukumar}(1985{\natexlab{a}})}]{onedDSUSY:1985}%
  \BibitemOpen
  \bibfield  {author} {\bibinfo {author} {\bibfnamefont {C~V}\ \bibnamefont
  {Sukumar}},\ }\bibfield  {title} {\enquote {\bibinfo {title} {Supersymmetric
  quantum mechanics of one-dimensional systems},}\ }\href {\doibase
  10.1088/0305-4470/18/15/020} {\bibfield  {journal} {\bibinfo  {journal}
  {Journal of Physics A: Mathematical and General}\ }\textbf {\bibinfo {volume}
  {18}},\ \bibinfo {pages} {2917--2936} (\bibinfo {year}
  {1985}{\natexlab{a}})}\BibitemShut {NoStop}%
\bibitem [{\citenamefont {Luo}\ \emph {et~al.}(2021)\citenamefont {Luo},
  \citenamefont {Raizen},\ and\ \citenamefont {Zhang}}]{Luo2021}%
  \BibitemOpen
  \bibfield  {author} {\bibinfo {author} {\bibfnamefont {Xi-Wang}\ \bibnamefont
  {Luo}}, \bibinfo {author} {\bibfnamefont {Mark~G.}\ \bibnamefont {Raizen}}, \
  and\ \bibinfo {author} {\bibfnamefont {Chuanwei}\ \bibnamefont {Zhang}},\
  }\bibfield  {title} {\enquote {\bibinfo {title} {Supersymmetry-assisted
  high-fidelity ground-state preparation of a single neutral atom in an optical
  tweezer},}\ }\href {\doibase 10.1103/PhysRevA.103.012415} {\bibfield
  {journal} {\bibinfo  {journal} {Phys. Rev. A}\ }\textbf {\bibinfo {volume}
  {103}},\ \bibinfo {pages} {012415} (\bibinfo {year} {2021})}\BibitemShut
  {NoStop}%
\bibitem [{\citenamefont {Heinrich}\ \emph {et~al.}(2014)\citenamefont
  {Heinrich}, \citenamefont {Miri}, \citenamefont {St{\"u}tzer}, \citenamefont
  {El-Ganainy}, \citenamefont {Nolte}, \citenamefont {Szameit},\ and\
  \citenamefont {Christodoulides}}]{Heinrich2014}%
  \BibitemOpen
  \bibfield  {author} {\bibinfo {author} {\bibfnamefont {Matthias}\
  \bibnamefont {Heinrich}}, \bibinfo {author} {\bibfnamefont {Mohammad-Ali}\
  \bibnamefont {Miri}}, \bibinfo {author} {\bibfnamefont {Simon}\ \bibnamefont
  {St{\"u}tzer}}, \bibinfo {author} {\bibfnamefont {Ramy}\ \bibnamefont
  {El-Ganainy}}, \bibinfo {author} {\bibfnamefont {Stefan}\ \bibnamefont
  {Nolte}}, \bibinfo {author} {\bibfnamefont {Alexander}\ \bibnamefont
  {Szameit}}, \ and\ \bibinfo {author} {\bibfnamefont {Demetrios~N.}\
  \bibnamefont {Christodoulides}},\ }\bibfield  {title} {\enquote {\bibinfo
  {title} {Supersymmetric mode converters},}\ }\href {\doibase
  10.1038/ncomms4698} {\bibfield  {journal} {\bibinfo  {journal} {Nature
  Communications}\ }\textbf {\bibinfo {volume} {5}},\ \bibinfo {pages} {3698}
  (\bibinfo {year} {2014})}\BibitemShut {NoStop}%
\bibitem [{\citenamefont {Infeld}\ and\ \citenamefont
  {Hull}(1951)}]{Hull:factorization:1951}%
  \BibitemOpen
  \bibfield  {author} {\bibinfo {author} {\bibfnamefont {L.}~\bibnamefont
  {Infeld}}\ and\ \bibinfo {author} {\bibfnamefont {T.~E.}\ \bibnamefont
  {Hull}},\ }\bibfield  {title} {\enquote {\bibinfo {title} {The factorization
  method},}\ }\href {\doibase 10.1103/RevModPhys.23.21} {\bibfield  {journal}
  {\bibinfo  {journal} {Rev. Mod. Phys.}\ }\textbf {\bibinfo {volume} {23}},\
  \bibinfo {pages} {21--68} (\bibinfo {year} {1951})}\BibitemShut {NoStop}%
\bibitem [{\citenamefont {Gutierrez}\ \emph {et~al.}(2018)\citenamefont
  {Gutierrez}, \citenamefont {Le{\'{o}}n}, \citenamefont {Belloni},\ and\
  \citenamefont {Robinett}}]{infinite_well_SUSY:2018}%
  \BibitemOpen
  \bibfield  {author} {\bibinfo {author} {\bibfnamefont {K}~\bibnamefont
  {Gutierrez}}, \bibinfo {author} {\bibfnamefont {E}~\bibnamefont
  {Le{\'{o}}n}}, \bibinfo {author} {\bibfnamefont {M}~\bibnamefont {Belloni}},
  \ and\ \bibinfo {author} {\bibfnamefont {R~W}\ \bibnamefont {Robinett}},\
  }\bibfield  {title} {\enquote {\bibinfo {title} {Exact results for the
  infinite supersymmetric extensions of the infinite square well},}\ }\href
  {\doibase 10.1088/1361-6404/aadc7f} {\bibfield  {journal} {\bibinfo
  {journal} {European Journal of Physics}\ }\textbf {\bibinfo {volume} {39}},\
  \bibinfo {pages} {065404} (\bibinfo {year} {2018})}\BibitemShut {NoStop}%
\bibitem [{\citenamefont {Cooper}\ \emph {et~al.}(1989)\citenamefont {Cooper},
  \citenamefont {Ginocchio},\ and\ \citenamefont
  {Wipf}}]{Cooper_1989:SUSY_Transform}%
  \BibitemOpen
  \bibfield  {author} {\bibinfo {author} {\bibfnamefont {F.}~\bibnamefont
  {Cooper}}, \bibinfo {author} {\bibfnamefont {J.~N.}\ \bibnamefont
  {Ginocchio}}, \ and\ \bibinfo {author} {\bibfnamefont {A.}~\bibnamefont
  {Wipf}},\ }\bibfield  {title} {\enquote {\bibinfo {title} {Supersymmetry,
  operator transformations and exactly solvable potentials},}\ }\href {\doibase
  10.1088/0305-4470/22/17/035} {\bibfield  {journal} {\bibinfo  {journal}
  {Journal of Physics A: Mathematical and General}\ }\textbf {\bibinfo {volume}
  {22}},\ \bibinfo {pages} {3707--3716} (\bibinfo {year} {1989})}\BibitemShut
  {NoStop}%
\bibitem [{\citenamefont {Lahrz}\ \emph {et~al.}(2017)\citenamefont {Lahrz},
  \citenamefont {Weitenberg},\ and\ \citenamefont
  {Mathey}}]{SUSY:Intertwine:2017}%
  \BibitemOpen
  \bibfield  {author} {\bibinfo {author} {\bibfnamefont {M.}~\bibnamefont
  {Lahrz}}, \bibinfo {author} {\bibfnamefont {C.}~\bibnamefont {Weitenberg}}, \
  and\ \bibinfo {author} {\bibfnamefont {L.}~\bibnamefont {Mathey}},\
  }\bibfield  {title} {\enquote {\bibinfo {title} {Implementing supersymmetric
  dynamics in ultracold-atom systems},}\ }\href {\doibase
  10.1103/PhysRevA.96.043624} {\bibfield  {journal} {\bibinfo  {journal} {Phys.
  Rev. A}\ }\textbf {\bibinfo {volume} {96}},\ \bibinfo {pages} {043624}
  (\bibinfo {year} {2017})}\BibitemShut {NoStop}%
\bibitem [{\citenamefont {Fernández~C.}\ and\ \citenamefont
  {Fernández‐García}(2004)}]{Hierarchy:2004}%
  \BibitemOpen
  \bibfield  {author} {\bibinfo {author} {\bibfnamefont {David~J.}\
  \bibnamefont {Fernández~C.}}\ and\ \bibinfo {author} {\bibfnamefont
  {Nicolás}\ \bibnamefont {Fernández‐García}},\ }\bibfield  {title}
  {\enquote {\bibinfo {title} {Higher‐order supersymmetric quantum
  mechanics},}\ }\href {\doibase 10.1063/1.1853203} {\bibfield  {journal}
  {\bibinfo  {journal} {AIP Conference Proceedings}\ }\textbf {\bibinfo
  {volume} {744}},\ \bibinfo {pages} {236--273} (\bibinfo {year}
  {2004})}\BibitemShut {NoStop}%
\bibitem [{\citenamefont {Liao}\ \emph {et~al.}(2010)\citenamefont {Liao},
  \citenamefont {Rittner}, \citenamefont {Paprotta}, \citenamefont {Li},
  \citenamefont {Partridge}, \citenamefont {Hulet}, \citenamefont {Baur},\ and\
  \citenamefont {Mueller}}]{Liao:10}%
  \BibitemOpen
  \bibfield  {author} {\bibinfo {author} {\bibfnamefont {Yean-an}\ \bibnamefont
  {Liao}}, \bibinfo {author} {\bibfnamefont {Ann Sophie~C.}\ \bibnamefont
  {Rittner}}, \bibinfo {author} {\bibfnamefont {Tobias}\ \bibnamefont
  {Paprotta}}, \bibinfo {author} {\bibfnamefont {Wenhui}\ \bibnamefont {Li}},
  \bibinfo {author} {\bibfnamefont {Guthrie~B.}\ \bibnamefont {Partridge}},
  \bibinfo {author} {\bibfnamefont {Randall~G.}\ \bibnamefont {Hulet}},
  \bibinfo {author} {\bibfnamefont {Stefan~K.}\ \bibnamefont {Baur}}, \ and\
  \bibinfo {author} {\bibfnamefont {Erich~J.}\ \bibnamefont {Mueller}},\
  }\bibfield  {title} {\enquote {\bibinfo {title} {Spin-imbalance in a
  one-dimensional {Fermi} gas},}\ }\href {\doibase 10.1038/nature09393}
  {\bibfield  {journal} {\bibinfo  {journal} {Nature}\ }\textbf {\bibinfo
  {volume} {467}},\ \bibinfo {pages} {567--569} (\bibinfo {year}
  {2010})}\BibitemShut {NoStop}%
\bibitem [{\citenamefont {Henderson}\ \emph {et~al.}(2009)\citenamefont
  {Henderson}, \citenamefont {Ryu}, \citenamefont {MacCormick},\ and\
  \citenamefont {Boshier}}]{Henderson_2009}%
  \BibitemOpen
  \bibfield  {author} {\bibinfo {author} {\bibfnamefont {K.}~\bibnamefont
  {Henderson}}, \bibinfo {author} {\bibfnamefont {C.}~\bibnamefont {Ryu}},
  \bibinfo {author} {\bibfnamefont {C.}~\bibnamefont {MacCormick}}, \ and\
  \bibinfo {author} {\bibfnamefont {M.~G.}\ \bibnamefont {Boshier}},\
  }\bibfield  {title} {\enquote {\bibinfo {title} {{Experimental demonstration
  of painting arbitrary and dynamic potentials for Bose{\textendash}Einstein
  condensates}},}\ }\href {\doibase 10.1088/1367-2630/11/4/043030} {\bibfield
  {journal} {\bibinfo  {journal} {New Journal of Physics}\ }\textbf {\bibinfo
  {volume} {11}},\ \bibinfo {pages} {043030} (\bibinfo {year}
  {2009})}\BibitemShut {NoStop}%
\bibitem [{\citenamefont {Barredo}\ \emph {et~al.}(2018)\citenamefont
  {Barredo}, \citenamefont {Lienhard}, \citenamefont {de~L{\'e}s{\'e}leuc},
  \citenamefont {Lahaye},\ and\ \citenamefont
  {Browaeys}}]{Paint_potential:SLM:2018}%
  \BibitemOpen
  \bibfield  {author} {\bibinfo {author} {\bibfnamefont {Daniel}\ \bibnamefont
  {Barredo}}, \bibinfo {author} {\bibfnamefont {Vincent}\ \bibnamefont
  {Lienhard}}, \bibinfo {author} {\bibfnamefont {Sylvain}\ \bibnamefont
  {de~L{\'e}s{\'e}leuc}}, \bibinfo {author} {\bibfnamefont {Thierry}\
  \bibnamefont {Lahaye}}, \ and\ \bibinfo {author} {\bibfnamefont {Antoine}\
  \bibnamefont {Browaeys}},\ }\bibfield  {title} {\enquote {\bibinfo {title}
  {Synthetic three-dimensional atomic structures assembled atom by atom},}\
  }\href {\doibase 10.1038/s41586-018-0450-2} {\bibfield  {journal} {\bibinfo
  {journal} {Nature}\ }\textbf {\bibinfo {volume} {561}},\ \bibinfo {pages}
  {79--82} (\bibinfo {year} {2018})}\BibitemShut {NoStop}%
\bibitem [{\citenamefont {Abanin}\ and\ \citenamefont
  {Levitov}(2005)}]{Levitov:2005}%
  \BibitemOpen
  \bibfield  {author} {\bibinfo {author} {\bibfnamefont {D.~A.}\ \bibnamefont
  {Abanin}}\ and\ \bibinfo {author} {\bibfnamefont {L.~S.}\ \bibnamefont
  {Levitov}},\ }\bibfield  {title} {\enquote {\bibinfo {title} {{Fermi-Edge
  Resonance and Tunneling in Nonequilibrium Electron Gas}},}\ }\href {\doibase
  10.1103/PhysRevLett.94.186803} {\bibfield  {journal} {\bibinfo  {journal}
  {Phys. Rev. Lett.}\ }\textbf {\bibinfo {volume} {94}},\ \bibinfo {pages}
  {186803} (\bibinfo {year} {2005})}\BibitemShut {NoStop}%
\bibitem [{\citenamefont {d'Ambrumenil}\ and\ \citenamefont
  {Muzykantskii}(2005)}]{Fermi_edge_res:2005}%
  \BibitemOpen
  \bibfield  {author} {\bibinfo {author} {\bibfnamefont {N.}~\bibnamefont
  {d'Ambrumenil}}\ and\ \bibinfo {author} {\bibfnamefont {B.}~\bibnamefont
  {Muzykantskii}},\ }\bibfield  {title} {\enquote {\bibinfo {title} {Fermi gas
  response to time-dependent perturbations},}\ }\href {\doibase
  10.1103/PhysRevB.71.045326} {\bibfield  {journal} {\bibinfo  {journal} {Phys.
  Rev. B}\ }\textbf {\bibinfo {volume} {71}},\ \bibinfo {pages} {045326}
  (\bibinfo {year} {2005})}\BibitemShut {NoStop}%
\bibitem [{\citenamefont {Knap}\ \emph
  {et~al.}(2012{\natexlab{b}})\citenamefont {Knap}, \citenamefont {Shashi},
  \citenamefont {Nishida}, \citenamefont {Imambekov}, \citenamefont {Abanin},\
  and\ \citenamefont {Demler}}]{Demler:ORTHO:2012}%
  \BibitemOpen
  \bibfield  {author} {\bibinfo {author} {\bibfnamefont {Michael}\ \bibnamefont
  {Knap}}, \bibinfo {author} {\bibfnamefont {Aditya}\ \bibnamefont {Shashi}},
  \bibinfo {author} {\bibfnamefont {Yusuke}\ \bibnamefont {Nishida}}, \bibinfo
  {author} {\bibfnamefont {Adilet}\ \bibnamefont {Imambekov}}, \bibinfo
  {author} {\bibfnamefont {Dmitry~A.}\ \bibnamefont {Abanin}}, \ and\ \bibinfo
  {author} {\bibfnamefont {Eugene}\ \bibnamefont {Demler}},\ }\bibfield
  {title} {\enquote {\bibinfo {title} {{Time-Dependent Impurity in Ultracold
  Fermions: Orthogonality Catastrophe and Beyond}},}\ }\href {\doibase
  10.1103/PhysRevX.2.041020} {\bibfield  {journal} {\bibinfo  {journal} {Phys.
  Rev. X}\ }\textbf {\bibinfo {volume} {2}},\ \bibinfo {pages} {041020}
  (\bibinfo {year} {2012}{\natexlab{b}})}\BibitemShut {NoStop}%
\bibitem [{\citenamefont {Schmidt}\ \emph {et~al.}(2018)\citenamefont
  {Schmidt}, \citenamefont {Knap}, \citenamefont {Ivanov}, \citenamefont {You},
  \citenamefont {Cetina},\ and\ \citenamefont {Demler}}]{KnapDemler:2018}%
  \BibitemOpen
  \bibfield  {author} {\bibinfo {author} {\bibfnamefont {Richard}\ \bibnamefont
  {Schmidt}}, \bibinfo {author} {\bibfnamefont {Michael}\ \bibnamefont {Knap}},
  \bibinfo {author} {\bibfnamefont {Dmitri~A}\ \bibnamefont {Ivanov}}, \bibinfo
  {author} {\bibfnamefont {Jhih-Shih}\ \bibnamefont {You}}, \bibinfo {author}
  {\bibfnamefont {Marko}\ \bibnamefont {Cetina}}, \ and\ \bibinfo {author}
  {\bibfnamefont {Eugene}\ \bibnamefont {Demler}},\ }\bibfield  {title}
  {\enquote {\bibinfo {title} {{Universal many-body response of heavy
  impurities coupled to a Fermi sea: a review of recent progress}},}\ }\href
  {\doibase 10.1088/1361-6633/aa9593} {\ \textbf {\bibinfo {volume} {81}},\
  \bibinfo {pages} {024401} (\bibinfo {year} {2018})}\BibitemShut {NoStop}%
\bibitem [{\citenamefont {del Campo}(2011)}]{Campo:decay_SP:2011}%
  \BibitemOpen
  \bibfield  {author} {\bibinfo {author} {\bibfnamefont {A.}~\bibnamefont {del
  Campo}},\ }\bibfield  {title} {\enquote {\bibinfo {title} {Long-time behavior
  of many-particle quantum decay},}\ }\href {\doibase
  10.1103/PhysRevA.84.012113} {\bibfield  {journal} {\bibinfo  {journal} {Phys.
  Rev. A}\ }\textbf {\bibinfo {volume} {84}},\ \bibinfo {pages} {012113}
  (\bibinfo {year} {2011})}\BibitemShut {NoStop}%
\bibitem [{\citenamefont {Goold}\ \emph {et~al.}(2011)\citenamefont {Goold},
  \citenamefont {Fogarty}, \citenamefont {Lo~Gullo}, \citenamefont
  {Paternostro},\ and\ \citenamefont {Busch}}]{Mossy:ORTHO:2011}%
  \BibitemOpen
  \bibfield  {author} {\bibinfo {author} {\bibfnamefont {J.}~\bibnamefont
  {Goold}}, \bibinfo {author} {\bibfnamefont {T.}~\bibnamefont {Fogarty}},
  \bibinfo {author} {\bibfnamefont {N.}~\bibnamefont {Lo~Gullo}}, \bibinfo
  {author} {\bibfnamefont {M.}~\bibnamefont {Paternostro}}, \ and\ \bibinfo
  {author} {\bibfnamefont {Th.}\ \bibnamefont {Busch}},\ }\bibfield  {title}
  {\enquote {\bibinfo {title} {{Orthogonality catastrophe as a consequence of
  qubit embedding in an ultracold Fermi gas}},}\ }\href {\doibase
  10.1103/PhysRevA.84.063632} {\bibfield  {journal} {\bibinfo  {journal} {Phys.
  Rev. A}\ }\textbf {\bibinfo {volume} {84}},\ \bibinfo {pages} {063632}
  (\bibinfo {year} {2011})}\BibitemShut {NoStop}%
\bibitem [{\citenamefont {Sukumar}(1985{\natexlab{b}})}]{Sukumar_1985}%
  \BibitemOpen
  \bibfield  {author} {\bibinfo {author} {\bibfnamefont {C.~V.}\ \bibnamefont
  {Sukumar}},\ }\bibfield  {title} {\enquote {\bibinfo {title} {{Supersymmetry,
  factorisation of the Schr\"{o}dinger equation and a Hamiltonian
  hierarchy}},}\ }\href {\doibase 10.1088/0305-4470/18/2/001} {\bibfield
  {journal} {\bibinfo  {journal} {Journal of Physics A: Mathematical and
  General}\ }\textbf {\bibinfo {volume} {18}},\ \bibinfo {pages} {L57--L61}
  (\bibinfo {year} {1985}{\natexlab{b}})}\BibitemShut {NoStop}%
\bibitem [{\citenamefont {Aronstein}\ and\ \citenamefont
  {Stroud}(1997)}]{Revivals:1997}%
  \BibitemOpen
  \bibfield  {author} {\bibinfo {author} {\bibfnamefont {David~L.}\
  \bibnamefont {Aronstein}}\ and\ \bibinfo {author} {\bibfnamefont {C.~R.}\
  \bibnamefont {Stroud}},\ }\bibfield  {title} {\enquote {\bibinfo {title}
  {Fractional wave-function revivals in the infinite square well},}\ }\href
  {\doibase 10.1103/PhysRevA.55.4526} {\bibfield  {journal} {\bibinfo
  {journal} {Phys. Rev. A}\ }\textbf {\bibinfo {volume} {55}},\ \bibinfo
  {pages} {4526--4537} (\bibinfo {year} {1997})}\BibitemShut {NoStop}%
\bibitem [{\citenamefont {Robinett}(2000)}]{Revivals:infinite_explained:2000}%
  \BibitemOpen
  \bibfield  {author} {\bibinfo {author} {\bibfnamefont {R.~W.}\ \bibnamefont
  {Robinett}},\ }\bibfield  {title} {\enquote {\bibinfo {title} {Visualizing
  the collapse and revival of wave packets in the infinite square well using
  expectation values},}\ }\href {\doibase 10.1119/1.19455} {\bibfield
  {journal} {\bibinfo  {journal} {American Journal of Physics}\ }\textbf
  {\bibinfo {volume} {68}},\ \bibinfo {pages} {410--420} (\bibinfo {year}
  {2000})}\BibitemShut {NoStop}%
\bibitem [{\citenamefont {Styer}(2001)}]{Revivals_revisited}%
  \BibitemOpen
  \bibfield  {author} {\bibinfo {author} {\bibfnamefont {Daniel~F.}\
  \bibnamefont {Styer}},\ }\bibfield  {title} {\enquote {\bibinfo {title}
  {Quantum revivals versus classical periodicity in the infinite square
  well},}\ }\href {\doibase 10.1119/1.1287355} {\bibfield  {journal} {\bibinfo
  {journal} {American Journal of Physics}\ }\textbf {\bibinfo {volume} {69}},\
  \bibinfo {pages} {56--62} (\bibinfo {year} {2001})}\BibitemShut {NoStop}%
\bibitem [{\citenamefont {Bluhm}\ \emph {et~al.}(1996)\citenamefont {Bluhm},
  \citenamefont {Kosteleck\'y},\ and\ \citenamefont
  {Porter}}]{Bluhm:WF_Revival:1996}%
  \BibitemOpen
  \bibfield  {author} {\bibinfo {author} {\bibfnamefont {Robert}\ \bibnamefont
  {Bluhm}}, \bibinfo {author} {\bibfnamefont {V.~Alan}\ \bibnamefont
  {Kosteleck\'y}}, \ and\ \bibinfo {author} {\bibfnamefont {James~A.}\
  \bibnamefont {Porter}},\ }\bibfield  {title} {\enquote {\bibinfo {title} {The
  evolution and revival structure of localized quantum wave packets},}\ }\href
  {\doibase 10.1119/1.18304} {\bibfield  {journal} {\bibinfo  {journal}
  {American Journal of Physics}\ }\textbf {\bibinfo {volume} {64}},\ \bibinfo
  {pages} {944--953} (\bibinfo {year} {1996})}\BibitemShut {NoStop}%
\bibitem [{\citenamefont {Robinett}(2004)}]{ROBINETT:Revivals:2004}%
  \BibitemOpen
  \bibfield  {author} {\bibinfo {author} {\bibfnamefont {R.W.}\ \bibnamefont
  {Robinett}},\ }\bibfield  {title} {\enquote {\bibinfo {title} {Quantum wave
  packet revivals},}\ }\href {\doibase
  https://doi.org/10.1016/j.physrep.2003.11.002} {\bibfield  {journal}
  {\bibinfo  {journal} {Physics Reports}\ }\textbf {\bibinfo {volume} {392}},\
  \bibinfo {pages} {1--119} (\bibinfo {year} {2004})}\BibitemShut {NoStop}%
\bibitem [{\citenamefont {Campisi}\ \emph {et~al.}(2011)\citenamefont
  {Campisi}, \citenamefont {H\"anggi},\ and\ \citenamefont
  {Talkner}}]{Fluc:2011}%
  \BibitemOpen
  \bibfield  {author} {\bibinfo {author} {\bibfnamefont {Michele}\ \bibnamefont
  {Campisi}}, \bibinfo {author} {\bibfnamefont {Peter}\ \bibnamefont
  {H\"anggi}}, \ and\ \bibinfo {author} {\bibfnamefont {Peter}\ \bibnamefont
  {Talkner}},\ }\bibfield  {title} {\enquote {\bibinfo {title} {{Colloquium:
  Quantum fluctuation relations: Foundations and applications}},}\ }\href
  {\doibase 10.1103/RevModPhys.83.771} {\bibfield  {journal} {\bibinfo
  {journal} {Rev. Mod. Phys.}\ }\textbf {\bibinfo {volume} {83}},\ \bibinfo
  {pages} {771--791} (\bibinfo {year} {2011})}\BibitemShut {NoStop}%
\bibitem [{\citenamefont {Stifter}\ \emph {et~al.}(1997)\citenamefont
  {Stifter}, \citenamefont {Leichtie}, \citenamefont {Schleich},\ and\
  \citenamefont {Marklof}}]{infinite_Talbot:1997}%
  \BibitemOpen
  \bibfield  {author} {\bibinfo {author} {\bibfnamefont {P.}~\bibnamefont
  {Stifter}}, \bibinfo {author} {\bibfnamefont {C.}~\bibnamefont {Leichtie}},
  \bibinfo {author} {\bibfnamefont {W.~P.}\ \bibnamefont {Schleich}}, \ and\
  \bibinfo {author} {\bibfnamefont {J.}~\bibnamefont {Marklof}},\ }\bibfield
  {title} {\enquote {\bibinfo {title} {{Das Teilchen im Kasten: Strukturen in
  der Wahrscheinlichkeitsdichte / The particle in a box : Structures in the
  probability}},}\ }\href {\doibase doi:10.1515/zna-1997-0501} {\bibfield
  {journal} {\bibinfo  {journal} {Zeitschrift für Naturforschung A}\ }\textbf
  {\bibinfo {volume} {52}},\ \bibinfo {pages} {377--385} (\bibinfo {year}
  {1997})}\BibitemShut {NoStop}%
\bibitem [{\citenamefont {Gro{\ss}mann}\ \emph {et~al.}(1997)\citenamefont
  {Gro{\ss}mann}, \citenamefont {Rost},\ and\ \citenamefont
  {Schleich}}]{SpaceT_infinte:1997}%
  \BibitemOpen
  \bibfield  {author} {\bibinfo {author} {\bibfnamefont {Frank}\ \bibnamefont
  {Gro{\ss}mann}}, \bibinfo {author} {\bibfnamefont {Jan-Michael}\ \bibnamefont
  {Rost}}, \ and\ \bibinfo {author} {\bibfnamefont {Wolfgang~P}\ \bibnamefont
  {Schleich}},\ }\bibfield  {title} {\enquote {\bibinfo {title} {Spacetime
  structures in simple quantum systems},}\ }\href {\doibase
  10.1088/0305-4470/30/9/004} {\bibfield  {journal} {\bibinfo  {journal}
  {Journal of Physics A: Mathematical and General}\ }\textbf {\bibinfo {volume}
  {30}},\ \bibinfo {pages} {L277--L283} (\bibinfo {year} {1997})}\BibitemShut
  {NoStop}%
\bibitem [{\citenamefont {Busch}\ \emph {et~al.}(1998)\citenamefont {Busch},
  \citenamefont {Anglin}, \citenamefont {Cirac},\ and\ \citenamefont
  {Zoller}}]{Busch:PB:1998}%
  \BibitemOpen
  \bibfield  {author} {\bibinfo {author} {\bibfnamefont {Th.}\ \bibnamefont
  {Busch}}, \bibinfo {author} {\bibfnamefont {J.~R.}\ \bibnamefont {Anglin}},
  \bibinfo {author} {\bibfnamefont {J.~I.}\ \bibnamefont {Cirac}}, \ and\
  \bibinfo {author} {\bibfnamefont {P.}~\bibnamefont {Zoller}},\ }\bibfield
  {title} {\enquote {\bibinfo {title} {{Inhibition of spontaneous emission in
  Fermi gases}},}\ }\href {\doibase 10.1209/epl/i1998-00426-2} {\bibfield
  {journal} {\bibinfo  {journal} {Europhysics Letters ({EPL})}\ }\textbf
  {\bibinfo {volume} {44}},\ \bibinfo {pages} {1--6} (\bibinfo {year}
  {1998})}\BibitemShut {NoStop}%
\bibitem [{\citenamefont {DeMarco}\ \emph {et~al.}(2001)\citenamefont
  {DeMarco}, \citenamefont {Papp},\ and\ \citenamefont {Jin}}]{PBlock:2001}%
  \BibitemOpen
  \bibfield  {author} {\bibinfo {author} {\bibfnamefont {B.}~\bibnamefont
  {DeMarco}}, \bibinfo {author} {\bibfnamefont {S.~B.}\ \bibnamefont {Papp}}, \
  and\ \bibinfo {author} {\bibfnamefont {D.~S.}\ \bibnamefont {Jin}},\
  }\bibfield  {title} {\enquote {\bibinfo {title} {{Pauli Blocking of
  Collisions in a Quantum Degenerate Atomic Fermi Gas}},}\ }\href {\doibase
  10.1103/PhysRevLett.86.5409} {\bibfield  {journal} {\bibinfo  {journal}
  {Phys. Rev. Lett.}\ }\textbf {\bibinfo {volume} {86}},\ \bibinfo {pages}
  {5409--5412} (\bibinfo {year} {2001})}\BibitemShut {NoStop}%
\bibitem [{\citenamefont {Mitchison}\ \emph {et~al.}(2020)\citenamefont
  {Mitchison}, \citenamefont {Fogarty}, \citenamefont {Guarnieri},
  \citenamefont {Campbell}, \citenamefont {Busch},\ and\ \citenamefont
  {Goold}}]{MOSSY:Thermo:2020}%
  \BibitemOpen
  \bibfield  {author} {\bibinfo {author} {\bibfnamefont {Mark~T.}\ \bibnamefont
  {Mitchison}}, \bibinfo {author} {\bibfnamefont {Thom\'as}\ \bibnamefont
  {Fogarty}}, \bibinfo {author} {\bibfnamefont {Giacomo}\ \bibnamefont
  {Guarnieri}}, \bibinfo {author} {\bibfnamefont {Steve}\ \bibnamefont
  {Campbell}}, \bibinfo {author} {\bibfnamefont {Thomas}\ \bibnamefont
  {Busch}}, \ and\ \bibinfo {author} {\bibfnamefont {John}\ \bibnamefont
  {Goold}},\ }\bibfield  {title} {\enquote {\bibinfo {title} {{In Situ
  Thermometry of a Cold Fermi Gas via Dephasing Impurities}},}\ }\href
  {\doibase 10.1103/PhysRevLett.125.080402} {\bibfield  {journal} {\bibinfo
  {journal} {Phys. Rev. Lett.}\ }\textbf {\bibinfo {volume} {125}},\ \bibinfo
  {pages} {080402} (\bibinfo {year} {2020})}\BibitemShut {NoStop}%
\end{thebibliography}%

\end{document}